\documentclass[a4paper,fleqn,usenatbib]{mnras}

\usepackage{newtxtext,newtxmath}

\usepackage[T1]{fontenc}
\usepackage{ae,aecompl}

\usepackage{graphicx}	
\usepackage{amsmath}	
\usepackage{amssymb}	
\usepackage[flushleft]{threeparttable}

\newcommand{\gsim}{\;\lower.6ex\hbox{$\sim$}\kern-7.75pt\raise.65ex\hbox{$>$}\;}
\newcommand{\lsim}{\;\lower.6ex\hbox{$\sim$}\kern-7.75pt\raise.65ex\hbox{$<$}\;}

\newcommand{\Myk}{M$_{\odot}$~yr$^{-1}$~kpc$^{-2}$}

\newcommand\Mhi{M_{\rm HI}}

\newcommand{\Msun}{M_{\odot}}

\newcommand\HII{\ion{H}{ii~}}
\newcommand\HI{\ion{H}{i~}}

\title[DDO~68: abundances and velocities]{Chemical abundances and radial velocities in the extremely metal-poor galaxy DDO~68 }

\author[F. Annibali et al.]{
F. Annibali,$^{1}$\thanks{E-mail: francesca.annibali@inaf.it}
V. La Torre,$^{2}$
M. Tosi,$^{1}$
C. Nipoti,$^{2}$
F. Cusano,$^{1}$
A. Aloisi,$^3$
\newauthor M. Bellazzini,$^{1}$ 
L. Ciotti,$^{2}$ 
A. Marchetti,$^{4}$
M. Mignoli,$^{1}$ 
 D. Romano,$^{1}$
E. Sacchi.$^3$
\\
$^{1}$INAF-Osservatorio di Astrofisica e Scienza dello Spazio, Via Gobetti 93/3, I-40129 Bologna, Italy\\
$^{2}$Dipartimento di Fisica e Astronomia, Bologna University, Via Gobetti 93/2, I-40129, Bologna, Italy\\
$^{3}$Space Telescope Science Institute, 3700 San Martin Drive, Baltimore, MD 21218\\
$^{4}$INAF-Istituto di Astrofisica Spaziale e Fisica Cosmica, Via Bassini 15, I-20133 Milano, Italy \\
}

\date{Accepted XXX. Received YYY; in original form ZZZ}

\pubyear{2018}

\begin{document}
\label{firstpage}
\pagerange{\pageref{firstpage}--\pageref{lastpage}}
\maketitle

\begin{abstract}

We present chemical abundances and radial velocities of six \HII regions in the extremely metal-poor star-forming dwarf galaxy DDO~68. They are derived from deep spectra in the wavelength range 3500 - 10,000 {\AA}, acquired with the Multi Object Double Spectrograph (MODS) at the Large Binocular Telescope (LBT).  In the three regions where the 
[O~III]$\lambda$4363  {\AA} line was detected, we inferred the abundance of He, N, O, Ne, Ar, and S through  the ``direct'' method. 
We also derived the oxygen abundances of all the six regions adopting indirect method calibrations. 
We confirm that DDO~68 is an extremely metal-poor galaxy, 
and a strong outlier in the luminosity - metallicity relation defined by star-forming galaxies. With the direct-method we find indeed an oxygen abundance  of 12+log(O/H)=7.14$\pm$0.07 in the northernmost region of the galaxy and, although with large uncertainties, an even lower 12+log(O/H)=6.96$\pm$0.09 in the ``tail''. This is, at face value, the most metal-poor direct abundance detection of any galaxy known. 
We derive a radial oxygen gradient of -0.06$\pm$0.03 dex/kpc (or -0.30 dex $R_{25}^{-1}$) with the direct method, and a steeper gradient of  -0.12$\pm$0.03 dex/kpc (or  -0.59  dex $R_{25}^{-1}$) from the indirect method.  
For the $\alpha$-element to oxygen ratios we obtain values in agreement with those found in other metal-poor star-forming dwarfs. For nitrogen, instead, we infer much higher values,  leading to log(N/O)$\sim -1.4$, at variance with the suggested existence of a tight plateau at $-1.6$ in extremely metal poor dwarfs. The derived helium mass fraction ranges from Y=0.240$\pm$0.005 to Y=0.25$\pm$0.02, compatible with standard big bang nucleosynthesis.
Finally, we measured \HII region radial velocities in the range 479$-$522 km/s from the tail to the head of the ``comet'', consistent with the rotation 
derived in the HI. 
 
\end{abstract}




\section{Introduction}

DDO~68 (UGC~5340, SDSS J0956+2849) is a star-forming dwarf galaxy of special interest in the framework of the $\Lambda$ Cold Dark Matter ($\Lambda$CDM) scenario for galaxy formation, because it is one of the first observed systems confirming the prediction that dark matter haloes host substructures down to the
resolution limit of the simulations \citep{Diemand08,Wheeler15}, i.e., down to the smallest mass scales.
In fact, in spite of being located in the huge Lynx-Cancer void
\citep{Pustilnik11} and of having formed a total mass of stars of only $10^8 M_{\sun}$  \citep{Sacchi16}, DDO~68 clearly suffers the presence of at least three smaller companions, thus appearing as {\it a flea with smaller fleas that on him prey}\footnote{From Jonathan Swift$'$s {\it On Poetry: a Rhapsody}: So, naturalists observe, a flea/ has smaller fleas that on him prey;/ and these have smaller still to bite 'em/ and so proceed ad infinitum.} \citep{anni16}.

DDO 68 is at a distance of $\sim$12.7 Mpc from us
\citep{Cannon14, Sacchi16,Makarov17}. With an oxygen abundance of 12+log(O/H)$\sim$7.14 measured in a few \HII regions at its northern edge \citep{Pustilnik05,Pustilnik07,izotov07a,Izotov09}, about 1/40 of
solar \citep[in][scale]{Caffau08}, it is an extremely
metal-poor (XMP) dwarf, much metal-poorer than
typical dwarfs of similar mass \citep{Pustilnik05}.  As a consequence, DDO~68 strongly deviates from the mean luminosity-metallicity relation defined by star-forming galaxies in the Local Universe \citep{lequeux79,skillman89,berg12}.

Because of its very distorted morphology, DDO 68 was suggested to be affected by galaxy
interaction already years ago \citep{Ekta08,Cannon14,Tikhonov14}. Ekta et al. noticed distortions in the \HI distribution and interpreted the observed features in terms of a late-stage merger of two gas-rich progenitors. 
Tikhonov et al. proposed that the portion of the galaxy that appears as a cometary tail is actually a
disrupted satellite currently being
cannibalized by DDO 68, while Cannon et al. identified a possible \HI
satellite, with $\Mhi\simeq 3\times 10^7 \Msun$, at a projected distance of $\simeq$40 kpc from the main body. \cite{anni16}, combining Hubble Space Telescope (HST) and Large Binocular Telescope (LBT) photometry, showed that, in addition to the known cometary tail, there is a faint, small stellar stream falling on the main body of DDO~68 and an arc possibly resulting from tidal interactions. They found that the morphologies of tail, stream, arc and main body are well reproduced by N-body simulations of the interaction among three distinct bodies, the main body and
two smaller ones with masses 1/10 and 1/150 the mass of the main body.

The faint stream is populated mainly by red giant branch (RGB) stars \citep{anni16}, while the cometary tail hosts stars of all ages \citep{Tikhonov14, Sacchi16,Makarov17} and is particularly rich in \HII regions. From deep and accurate Colour-Magnitude diagrams (CMDs) obtained from HST photometry, \cite{Sacchi16} inferred a robust spatially resolved star formation history (SFH) of the various regions of DDO~68, finding that everywhere stars were already forming at the oldest lookback time reached by the photometry, i.e. at least 2-3 Gyr ago, corresponding to the minimum age of the measured RGB stars. The measured RGB stars may actually be up to 13 Gyr old, but the age-metallicity degeneracy affecting the RGB prevents one to precisely age-date these stars, since independent spectroscopic estimates of their metallicity are not available. According to Sacchi et al., DDO~68 is likely to have formed stars at a low rate during its whole life, with a significant enhancement a few hundred million years ago, a peak activity of $\sim4.6 \times 10^{-2}$  \Myk ~ between 20 and 60 Myr ago, and a current much lower rate.

If the galaxy has formed stars for a Hubble time, although at a relatively low rate, how did it manage to remain so metal-poor? Is it because it has recently accreted significant amounts of very low-metallicity gas that diluted its medium, or because it got rid through  differential galactic winds of the elements produced by its stars? Chemical evolution models of late-type dwarfs have often invoked both effects to reproduce the observed low chemical abundances of star-forming dwarfs \citep[see e.g.,][]{mt85,pilyugin92,marconi94,romano06,recchi08,romano13}, suggesting that the outflows triggered by supernova explosions in these shallow potential well systems 
 may be more efficient in ejecting some of the heavy elements (like the $\alpha$-elements and iron) than other chemical species such as H and He. 
Indeed, hydrodynamical simulations of late-type dwarfs  \citep[e.g.,][]{derco99,maclow99,scan10} clearly show that the outflows that blow out of these galaxies drive substantial amounts of the metals produced in the starburst into the intergalactic medium. 
Other authors have instead considered the infall of primordial gas sufficient \citep[see, e.g.][]{gavilan13} to reproduce the observed metallicities.
It has also been suggested  \citep[e.g.][]{legrand00} that a continuous, very low star formation regime could account for the observed abundances in some extremely metal poor dwarf galaxies.

The combination of a detailed SFH with chemical abundances derived from accurate spectroscopy and with constraints on the dynamics of its various constituents is a key ingredient to reconstruct a coherent and complete picture of how DDO~68 (and, more in general, XMPs) formed and evolved.  
To this aim, here we exploit the Multi Object Double Spectrograph (MODS) mounted on the LBT to study DDO~68's \HII regions. 

In this paper, Section~2 describes the observations and data reduction, and Section~3 presents the procedure for the derivation of the reddening-corrected emission line fluxes. The results of temperatures, densities, chemical abundances are described in Section~4, while the derived abundances, abundance ratios and abundance spatial distributions are described in Section~5. Radial velocities are presented and discussed in Section 6. The overall results are discussed and summarised in Section~7.

\section{Observations and data reduction \label{data_reduction}}

\begin{figure}
\includegraphics[width=\columnwidth]{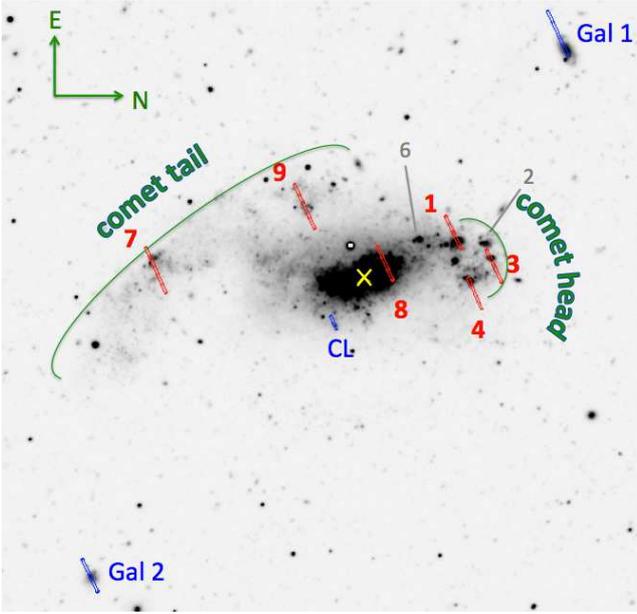}
 \caption{LBT $g$ image of DDO~68, showing the position of our 9 MODS slits. Six slits (1, 3, 4, 7, 8 and 9) were centred on \HII regions in DDO~68,  two slits (Gal~1 and Gal~2) on two candidate galaxy companions, and one slit (dubbed CL) on a candidate star cluster in DDO~68.  For completeness, we also indicate (in grey) Reg.~2 and Reg.~6 identified  by \protect  \cite{Pustilnik05} and not observed in our study because of slit positioning constraints.
 The head and the tail of the ``comet'' are indicated on the image.
 The yellow cross corresponds to the centre of the galaxy identified from isophotal contours derived from the LBT $g$ image.
}
\label{image}
\end{figure}

Our observations (LBT program 2016-2017-50, PI Annibali) were performed on February 21, 2017, with MODS1 and MODS2 mounted on LBT.
 The LBT, located at Mt. Graham  (Arizona), is composed of  two identical 8.4 m telescopes that can work simultaneously in binocular mode, and it is larger than all telescopes used for previous optical spectroscopic studies of DDO~68. When used in dichroic mode, the MODS spectrographs provide high quality spectra in the 
$\sim$3400-5500 \AA \ and $\sim$5700-9800 \AA \ ranges for the blue and red channels, respectively. 
The 6$^{\prime}\times 6^{\prime}$ field of view of MODS is perfectly suited to cover the whole DDO~68 system and its surroundings. 
The target \HII regions were selected from our HST (GO program 11578; PI: Aloisi) V, I, and H$\alpha$  images.  In those images, \HII regions are resolved and appear as regions of diffuse H$\alpha$ and V (i.e. [OIII]$\lambda\lambda$4959,5007)  emission. 
In order to look for differences (or lack thereof) in the abundances of the different structures, we selected \HII regions covering all the main structural zones of DDO~68, and targeted six of them for our observations. 
According to H$\alpha$ images from our HST program and from \cite{Pustilnik05}, no emission line region is found along the faint stream discovered by \cite{anni16}. In addition, we targeted a candidate star cluster (CL) and two potential companions of DDO~68 detected in our LBT images (dubbed Gal~1 and Gal~2).  An LBT image of DDO~68 with overimposed the positions of our 9 slits is shown in Figure~\ref{image}. 

Four of our target \HII regions were already observed by \cite{Pustilnik05} and \cite{Izotov09}; for them we adopt the names assigned by those authors, i.e. regions 1, 3, 4, and 7 (see Fig.~\ref{image}.) Regions 8 and 9  are instead newly observed \HII regions. 
One \HII region (Reg.~8) is close to the centre of the main body, two (Reg.~7 and Reg.~9) lie along the cometary tail, and three (Reg.~1, Reg.~3 and Reg.~4) in the northern ring-like structure that some consider the head of the comet \citep{Makarov17}. Reg.~3 also includes the Luminous Blue Variable (LBV) studied by \cite{Izotov09} and \cite{Pustilnik16}

Observations and data reduction have been designed following the same strategy devised by \cite{annibali17} for the star-forming irregular galaxy NGC~4449. We have crafted the masks using our HST images, and adopting a slit width of 1$^{\prime\prime}$; for the H~II regions, we adopted slit lengths between 15'' and 20'', sufficient to include also a portion of the background without significant nebular emission.
We designed two different masks, implemented in two separate runs (Run A and Run B) to be able to place the slits on all our 9 target regions. 
We observed our targets using the blue G400L (3200$-$5800 \AA) and the red G670L (5000$-$10000 \AA) gratings on the blue and red channels in dichroic mode, to cover the spectral range from the [OII]$\lambda$3727 nebular line to the [SIII]$\lambda$9530 line, allowing the derivation of the abundances of He, N, O, Ne, S, and Ar.  
The observations were performed in binocular mode using MODS1 and MODS2 simultaneously. 
The exposure times were dictated by the need to observe the relatively faint [OIII]$\lambda$4363 auroral line with a signal/noise of at least 3, to derive the nebular electronic temperatures. We had 12 exposures in Run A and 10 in Run B of 600 sec each, for a total of less than 4 hours.   The seeing varied between $\sim$0.5'' and $\sim$0.8'', 
and the airmass from $\sim$1.0 to   $\sim$1.1. In order to avoid significant effects from differential atmospheric refraction \citep[see e.g.][]{filippenko82}, the slits were 
approximately oriented along the parallactic angle (see Table~\ref{journal}). 
We used the MODS observing planning tool\footnote{www.astronomy.ohio-state.edu/MODS/ObsTools/modsTools/} to check that our observations were not affected by flux losses due to differential atmospheric refraction indeed.

 \begin{table*}
  \caption{Journal of Observations.}
  \label{journal}
  \begin{tabular}{lccccccccc}
\hline
Run & Exp. n. & Instrument & Date-obs. & Time-obs. & Exp.Time & Seeing & Airmass & PA & PARA \\
&. &  &[yyyy-mm-dd] & [hh:mm:ss] & [s] & [arcsec] &  & [$^{\circ}$] &  [$^{\circ}$]  \\

 \hline         
& 1  &  MODS1 & 2017-02-21 & 05:22:48 &  600.  &  0.6''-0.7'' &  1.09  & -65  &   -74 \\
& 2  &  MODS2 & 2017-02-21 & 05:22:28 &  600.  &  0.6''-0.7'' &  1.09  & -65  &   -74 \\
& 3  &  MODS1 & 2017-02-21 & 05:34:32 &  600.  &  0.6''-0.7'' &  1.07  & -65  &   -73 \\
& 4  &  MODS2 & 2017-02-21 & 05:34:16 &  600.  &  0.6''-0.7'' &  1.07  & -65  &   -73 \\
B & 5  &  MODS1 & 2017-02-21 & 05:46:16 &  600.  &  0.6''-0.7'' &  1.06  & -65  &   -73 \\
& 6  &  MODS2 & 2017-02-21 & 05:47:12 &  600.  &  0.6''-0.7'' &  1.06  & -65  &   -73 \\
& 7  &  MODS1 & 2017-02-21 & 05:58:00 &  600.  &  0.6''-0.7'' &  1.04  & -65  &   -72 \\
& 8  &  MODS2 & 2017-02-21 & 05:57:60 &  600.  &  0.5''-0.6'' &  1.04  & -65  &   -72 \\
& 9  &  MODS1 & 2017-02-21 & 06:09:44 &  600.  &  0.5''-0.6'' &  1.03  & -65  &   -70 \\
& 10 &  MODS2 & 2017-02-21 & 06:10:47 &  600.  &  0.5''-0.6'' &  1.03  & -65  &   -70 \\
\hline \\
& 1  &   MODS1 & 2017-02-21 & 06:38:02 & 600. &  0.5''-0.6'' &  1.01 & -65  &  -59 \\
& 2  &   MODS2 & 2017-02-21 & 06:37:40 & 600. &  0.5''-0.6'' &  1.01 & -65  &  -59 \\
& 3  &   MODS1 & 2017-02-21 & 06:49:47 & 600. &  0.5''-0.6'' &  1.01 & -65  &  -48 \\
& 4  &   MODS2 & 2017-02-21 & 06:49:27 & 600. &  0.5''-0.6'' &  1.01 & -65  &  -48 \\
& 5  &   MODS1 & 2017-02-21 & 07:01:31 & 600. &  0.5''-0.6'' &  1.00   & -65  &  -27 \\
A & 6  &   MODS2 & 2017-02-21 & 07:01:15 & 600. &  0.5''-0.6'' &  1.00   & -65  &  -27 \\
& 7  &   MODS1 & 2017-02-21 & 07:13:15 & 600. &  0.5''-0.6'' &  1.00   & -65  &  8 \\
& 8  &   MODS2 & 2017-02-21 & 07:13:03 & 600. &  0.5''-0.6'' &  1.00   & -65  &  8\\
& 9  &   MODS1 & 2017-02-21 & 07:24:59 & 600. &  0.5''-0.6'' &  1.00  & -65  &   37\\
& 10  &  MODS2 & 2017-02-21 & 07:25:50 & 600. &  0.5''-0.6'' &  1.00  & -65  &  37\\
& 11  &  MODS1 & 2017-02-21 & 07:36:43 & 600. &  0.8''       &  1.01 & -65  &  54\\
& 12  &  MODS2 & 2017-02-21 & 07:36:38 & 600. &  0.8''       &  1.01 & -65  &  54\\
\hline
\hline
 \end{tabular}
 \begin{tablenotes}
\small
\item Col.~(1): run of observations; Col.~(2): exposure number; Col.~(3): instrument (MODS1 or MODS2); Col.~(4): date of observations;  Col.~(5): UTC time of observations; 
  Col.~(6): duration of exposure;  Col.~(7): seeing in arcsec; Col.~(8): airmass;  Col.~(9): slit position angle;  Col.~(10): parallactic angle. 
    \end{tablenotes}
 \end{table*}

The data acquired with MODS1 and MODS2 were treated separately, since the two instruments turned out to have different sensitivities. 
However, to allow for the potential detection of intrinsically faint, key lines (e.g., [O~III]$\lambda$4363), we also produced deep spectra combining together all the MODS1 and MODS2 sub-exposures. Bias and flat-field corrections, and wavelength calibration were executed  by the Italian LBT Spectroscopic Reduction Facility at INAF-IASF Milano. They provided
the calibrated two-dimensional (2D) spectra for the individual sub-exposures. 
The accuracy of the wavelength calibration obtained from the pipeline was checked against prominent sky lines adopting the nominal sky wavelengths tabulated in \citet{uves_sky}. 
Then, the individual sub-exposures were sky-subtracted and combined into 
final 2D blue and red frames. 
The sky subtraction was performed with the {\it{background}} task in IRAF\footnote{ IRAF is distributed by the National Optical Astronomy Observatory, which is operated by the Association of Universities for Research in Astronomy, Inc., under cooperative agreement with the National Science Foundation.}, typically choosing the windows at the two opposite sides of the central source. 
The typical size of our H~II regions (FWHM$\lesssim$2'') is significantly smaller than the adopted slit lengths of 15''-20'', allowing for an adequate removal of the sky background.
The  one-dimensional (1D) spectra were extracted from the 2D wavelength-calibrated and sky-subtracted spectra by running the {\it apall} task in the {\it twodspec} IRAF package. 
The effective spectral resolution of our spectra, as measured from prominent sky lines in the 1D background-unsubtracted spectra, is R$\sim$1400 at $\lambda\sim5500$\ \AA~ and 
R$\sim$1000 at $\lambda\sim7200$\ \AA. 

The blue and red 1D spectra were flux calibrated using the sensitivity curves from the Italian LBT spectroscopic reduction pipeline; the curves  were derived using the spectrophotometric standard star GD~71, observed with both MODS1 and MODS2 in dichroic mode with a 5$^{\prime\prime}$-width slit on the same night as our targets and at the same airmass. 
Each MODS1/MODS2 spectrum was calibrated using the appropriate sensitivity functions, while the deep, total combined spectra were calibrated through an average sensitivity curve. After calibration, the MODS1, MODS2, and total combined spectra turned out to typically agree within 5$-$10~\%.   
The final calibrated MODS1 and MODS2 spectra of our nine targets are shown in Figure~\ref{spectra_reg1} and Figures~\ref{spectra_reg3} to ~\ref{spectra_cl} in the Appendix. 
As shown by the spectra, Gal.~1 and Gal.~2 turned out to be background emission-line galaxies at redshifts of z$\sim$0.11 and  z$\sim$0.0148, respectively, and CL an 
early-type galaxy at  z$\sim$0.42. Since they are not related to DDO~68, these systems will be no longer discussed in the paper.

\begin{figure*}
\includegraphics[width=\textwidth]{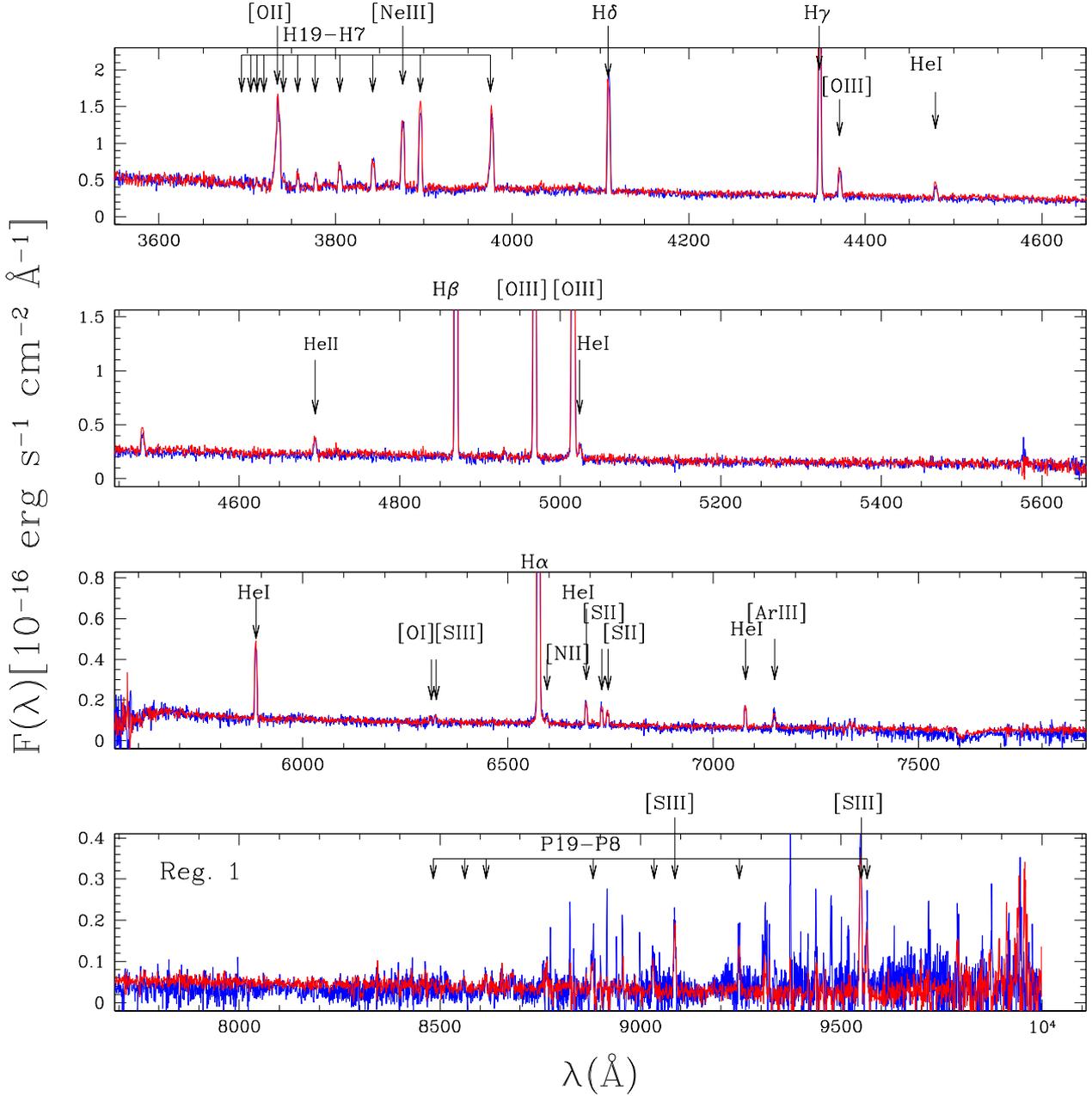}
 \caption{LBT/MODS spectra for Reg.~1 in  DDO~68 acquired with MODS1 (blue line) and MODS2 (red line). Indicated are the main identified emission lines. The spectra of the other \HII regions studied in this paper  (Reg.~3, 4, 7, 8 and 9), of the two candidate companion galaxies (Gal.~1 and Gal.~2), and of the candidate star cluster (CL) are shown in the Appendix.
 }
\label{spectra_reg1}
\end{figure*}

\section{Emission line measurement} \label {section_fluxes}

Emission line fluxes were measured with the {\it deblend} function in the  {\it splot} IRAF task. 
We fitted all together lines close in wavelengths (e.g. H$\beta$, HeI$\lambda$4922, [O~III]$\lambda\lambda4959,5007$, HeI$\lambda$5015) assuming a Gaussian profile, leaving the centroids and FWHM as free parameters, but imposing  the FWHMs to be the same for all fitted lines. 
The continuum was defined choosing two continuum windows to the left and to the right of the line or line complex, and fitting it with a linear regression. 
In the case of the [N~II]$\lambda\lambda$6548,84 lines, blended with the much stronger H$\alpha$ emission and therefore difficult to measure, we did not leave the    centroids as free but allowed a single shift with respect to the nominal zero-velocity line wavelengths. Furthermore,  we introduced an additional broad H$\alpha$ component in the fit of Reg. 1 and 3 to properly reproduce the observed wings.
The final emission fluxes were obtained repeating the measurement several times with slightly different continuum choices, and averaging the results. The procedure was applied separately to the MODS1 and the MODS2 spectra. We found that in general MODS2 provided fluxes about 5$\%$ higher than MODS1.
The final line fluxes were computed averaging the results from MODS1 and MODS2, and the errors were computed as the standard deviation from the two different measurements. 
The detection of intrinsically faint lines that are key to the determination of the electronic temperatures (e.g.,  [O~III]$\lambda$4363, [S~III]$\lambda$6312, [O~II]$\lambda\lambda$7320,30) was checked on the 2D spectra. The  [O~III]$\lambda$4363 line is well detected in Reg.~1 and Reg.~3, is quite faint in Reg.~7, and is not detected in the other regions. 
For Reg.~7, the measurement of this important feature was performed on the deep total combined spectrum. 

Reg.~3 is known to host a LBV, discovered by \cite{Pustilnik08} from the presence of broad emission components, P-Cygni profiles, and spectral variability. 
The LBV experienced a very rare event of a giant eruption during the years 2008-2011, and was in a ``faint'' state during the years 2015-2016 \citep{Pustilnik17}.  
During the outburst event, the H$\beta$ emission of Reg. 3 was observed to overcome that of the [O~III]$\lambda$5007 line, as shown in Fig.~1 of \cite{Pustilnik08}. All these properties are not observed in our data of Reg. 3, which look more simular to the spectrum  acquired in 2005 by Pustilnik et al, when the LBV was at its minimum. We conclude therefore that at the epoch of our observations the LBV was, as during the years 2015-2016, in a rather quiescent phase.

The derived emission lines for the six \HII regions studied in DDO~68 are provided in Table~\ref{tab_flux_raw} in the Appendix.

Balmer underlying absorption and reddening were derived simultaneously through a $\chi^2$ minimization procedure. Based on the measured 
H$\delta$, H$\gamma$, H$\beta$ and H$\alpha$ fluxes, we constructed the following  $\chi^2$ from all possible line combinations:

\begin{align}
 \nonumber \chi^2 = \frac{(R_{\alpha,\beta}^{corr} -  R_{\alpha,\beta}^{teo})^2}{\sigma_{\alpha,\beta}^2} + 
 \frac{(R_{\alpha,\gamma}^{corr} -  R_{\alpha,\gamma}^{teo})^2}{\sigma_{\alpha,\gamma}^2} + 
\frac{(R_{\alpha,\delta}^{corr} -  R_{\alpha,\delta}^{teo})^2}{\sigma_{\alpha,\delta}^2} + \\
\frac{(R_{\beta,\gamma}^{corr} -  R_{\beta,\gamma}^{teo})^2}{\sigma_{\beta,\gamma}^2} + 
 \frac{(R_{\beta,\delta}^{corr} -  R_{\beta,\delta}^{teo})^2}{\sigma_{\beta,\delta}^2} + 
 \frac{(R_{\gamma,\delta}^{corr} -  R_{\gamma,\delta}^{teo})^2}{\sigma_{\gamma,\delta}^2}, 
 \label{eq1}
  \end{align}

\noindent where, for instance, $R_{\alpha,\beta}^{corr}$ is the reddening-corrected H$\alpha$ to H$\beta$ ratio:   

\begin{align}
\nonumber  R_{\alpha,\beta}^{corr} = \frac{F'_{H\alpha}}{F'_{H\beta}} \times 10^{0.4(A_{H\alpha} - A_{H\beta})}\\
= \frac{F'_{H\alpha}}{F'_{H\beta}} \times 10^{0.4 \left( \frac{A_{H\alpha}}{A_V} - \frac{A_{H\beta}}{A_V} \right)\times R_V \times E(B-V)} 
 \label{eq2}
  \end{align}

\noindent where $A_{H\alpha}/A_V$ and $A_{H\beta}/A_V$ are taken from the \cite{cardelli89} extinction law with $R_V = 3.05$,  $E(B-V)$ is the unknown reddening, 
and $F'_{H\alpha}$ and $F'_{H\beta}$ are the emission fluxes corrected for underlying Balmer absorption according to the formula:

\begin{equation}
 F'_{H\alpha,\beta} = F_{H\alpha,\beta} + EW_{abs} \times C_{H\alpha,\beta}.
 \label{eq3}
\end{equation}

In Eq.~\ref{eq3},  $F_{H\beta}$  and $F_{H\alpha}$ are the emission fluxes measured on the spectra,  $C_{H\beta}$ and   $C_{H\alpha}$ are the local continua, and 
$EW_{abs}$ is the underlying unknown Balmer absorption in equivalent width assumed to be the same for H$\delta$, H$\gamma$,  H$\beta$, and  H$\alpha$. This is a reasonable assumption since simple stellar populations models \citep[e.g.][hereafter SB99]{sb99} show that the equivalent width ratios of these absorption lines are close to unity. 
As for the theoretical Balmer line ratios, we adopted values of  $R_{\alpha,\beta}^{teo}=2.79$,   $R_{\alpha,\gamma}^{teo}=5.90$, $R_{\alpha,\delta}^{teo}=10.60$,
$R_{\beta,\gamma}^{teo}=2.12$, $R_{\beta,\delta}^{teo}=3.80$, $R_{\beta,\delta}^{teo}=1.80$ assuming case B recombination with $T_e=20,000$ K and $n_e=100 \ cm^{-3}$
\citep{oster89,storey95}, typical of very-low metallicity \HII regions. Finally, the $\sigma$ uncertainties in  Eq. (1) are obtained by propagating the flux errors on the emission line ratios. 

The $\chi^2$ minimization provides reddening values in the range E(B-V)$\sim0.07-0.24$ mag and absorption corrections below EW$\sim$1.5 \AA \ for our six \HII regions in DDO~68, consistent with previous works \citep[e.g.][]{Pustilnik05}.  
As for Helium lines, the absorption correction was directly estimated from SB99 simple stellar population models with Z$=$0.001 (the lowest metallicity available for the high-resolution spectra products) and age$<$10 Myr, resulting into EW(He~I$\lambda$4471)$\sim$0.5 \AA, EW(He~I$\lambda$5876)$\sim$0.5 \AA, and EW(He~I$\lambda$6678)$\sim$0.3 \AA.  The assumption of a young SSP for the underlying continuum is a reasonable choice since, although DDO~68 experienced a quite continuous SFH from several Gyrs ago until present \citep{Sacchi16}, the integrated stellar light within its H~II regions is dominated by young massive stars.
The derived E(B$-V$) values and the final emission-line fluxes corrected for reddening and underlying absorption are provided in Table~\ref{tab_flux_corr} for our six \HII regions in DDO~68.

 \begin{table*}
  \caption{Reddening-corrected Emission Fluxes for  \HII regions in DDO~68.}
  \label{tab_flux_corr}
  \begin{tabular}{lcccccc}
\hline
Line & Reg-1 & Reg-3 & Reg-4 & Reg-7 & Reg-8 & Reg-9 \\
 \hline         
 {[O II]} $\lambda$3727 &  27 $\pm$  1 &  63 $\pm$  5 &  84 $\pm$  2 &  51 $\pm$  7 & 290 $\pm$ 20 & 150 $\pm$ 20 \\
H10 $\lambda$3978 &    4.5 $\pm$   0.2 &    4.0 $\pm$   0.6 &  $-$ &    5.9 $\pm$   0.5 &  $-$ &  $-$ \\
He I $\lambda$3820  & $-$ &   2.8 $\pm$  0.2 & $-$ & $-$ & $-$ & $-$ \\
H9$+$He II $\lambda$3835 &    6.2 $\pm$   0.3 &    5.8 $\pm$   0.6 &  $-$ &    5 $\pm$  2 &  $-$ &  $-$ \\
{[Ne III]} $\lambda$3869 &   15.6 $\pm$   0.5 &   12.3 $\pm$   0.8 &   10 $\pm$   1 &    7.3 $\pm$   0.2 &  $-$ &  $-$ \\
H8$+$He I $\lambda$3889 &   19 $\pm$   1 &   17 $\pm$   2 &   13.2 $\pm$   0.7 &   18 $\pm$   2 &   18 $\pm$   2 &  $-$ \\
H$\epsilon$ $+$ He I $+$[Ne III] $\lambda$3970 &   19.1 $\pm$   0.7 &   18 $\pm$   1 &   12 $\pm$   1 &   16.6 $\pm$   0.8 &   16 $\pm$   2 &  $-$ \\
H$\delta$ $\lambda$4101 &   26.4 $\pm$   0.9 &   27 $\pm$   2 &   26.4 $\pm$   0.2 &   26.3 $\pm$   0.4 &   26 $\pm$   2 &   32 $\pm$   3 \\
H$\gamma$ $\lambda$4340 &   48 $\pm$   2 &   48 $\pm$   4 &   47.7 $\pm$   0.6 &   47.6 $\pm$   0.9 &   45 $\pm$  2 &   50$\pm$   8 \\
{[O III]} $\lambda$4363 &   5.8 $\pm$  0.2 &   3.5 $\pm$  0.6 &  $-$ &   1.7 $\pm$  0.1 &   $-$ & $-$ \\
He I $\lambda$4471 &    3.7 $\pm$   0.3 &  $-$ &    4.0 $\pm$   0.2 &    4.6 $\pm$   0.1 &  $-$ &  $-$ \\
He II (WR) $\lambda$4686 & $-$ &   8 $\pm$  2 & $-$ & $-$ & $-$ & $-$ \\
He II  $\lambda$4686 &   2.30 $\pm$  0.07 & $-$ & $-$ & $-$ & $-$ & $-$ \\
H$\beta$ $\lambda$4861 &  100 $\pm$   4&  100 $\pm$   7 &  100 $\pm$   1 &  100 $\pm$   2 &  100 $\pm$   5 &  100 $\pm$   1 \\
He I $\lambda$4922 &   1.03 $\pm$  0.07 & $-$ & $-$ & $-$ & $-$ & $-$ \\
{[O III]} $\lambda$4959 &   63 $\pm$   3 &   43 $\pm$  3 &   25.4 $\pm$   0.9 &   23.5 $\pm$   0.4 &   20 $\pm$   1 &   13 $\pm$   1 \\
{[O III]} $\lambda$5007 &  181 $\pm$   8 &  125 $\pm$  9 &   73 $\pm$   2 &   66 $\pm$   1 &   61 $\pm$  4 &   50 $\pm$   6 \\
He I $\lambda$5015  &   2.3 $\pm$  0.1 & $-$ & $-$ & $-$ & $-$ & $-$ \\
He I $\lambda$5876  &    9.3 $\pm$   0.3 &    9.8 $\pm$   0.5 &    7.8 $\pm$   0.2 &    9.5 $\pm$   0.2 &    9 $\pm$  2 &  $-$ \\
{[OI]}  $\lambda$6302 &   0.50 $\pm$  0.04 & $-$ & $-$ &   1.4 $\pm$  0.4 &   6.3 $\pm$  0.5 & $-$ \\
{[S III]} $\lambda$6312 &   0.67 $\pm$  0.05 & $-$ & $-$ & $-$ & $-$ & $-$ \\
H$\alpha$ $\lambda$6563  &  281 $\pm$  10 &  290 $\pm$  20 &  281 $\pm$   3 &  280 $\pm$   4 &  270 $\pm$ 10 &  300 $\pm$  30 \\
{[N II]} $\lambda$6584  &    1.0 $\pm$   0.2 &    2.3 $\pm$   0.1 &    2.7 $\pm$   0.4 &    1.9 $\pm$   0.1 &    8.0 $\pm$   0.4 &  $-$ \\
He I $\lambda$6678   &    2.7 $\pm$   0.2 &    2.8 $\pm$   0.2 &    2.8 $\pm$   0.1 &    2.9 $\pm$   0.2 &    2.5 $\pm$   0.3 &  $-$ \\
{[S II]} $\lambda$6716  &    2.1 $\pm$   0.1 &    4.8 $\pm$   0.6 &    8 $\pm$  3 &    4.9 $\pm$   0.1 &   24 $\pm$   1 &   14 $\pm$  2 \\
{[S II]} $\lambda$6731  &    1.5 $\pm$   0.1 &    3.6 $\pm$   0.2 &    3.8 $\pm$   0.8 &    3.7 $\pm$   0.4 &   18 $\pm$   1 &    6 $\pm$   1 \\
He I $\lambda$7065  &    2.3 $\pm$   0.1 &    2.5 $\pm$   0.2 &    2.4 $\pm$   0.2 &    1.8 $\pm$   0.1 &  $-$ &  $-$ \\
{[Ar III]} $\lambda$7136  &    1.7 $\pm$   0.1 &    1.9 $\pm$   0.1 &    1.5 $\pm$   0.1 &    1.4 $\pm$   0.4 &  $-$ &  $-$ \\
He I $\lambda$7281  &   0.42 $\pm$  0.02 & $-$ & $-$ & $-$ & $-$ & $-$ \\
{[O II]} $\lambda$7320  &   0.67 $\pm$  0.08 & $-$ & $-$ & $-$ &   3.2 $\pm$  0.3 & $-$ \\
{[O II]} $\lambda$7330  &   0.5 $\pm$  0.1 & $-$ & $-$ & $-$ &   3.9 $\pm$  0.6 & $-$ \\
P10 $\lambda$9017  &   21.3 $\pm$   0.7 &  $-$ &  $-$ &  $-$ &  $-$ &  $-$ \\
{[S III]} $\lambda$9069  &    3.5 $\pm$   0.2 &    3.2 $\pm$   0.3 &    3 $\pm$   1 &    2.4 $\pm$   0.7 &    3.5 $\pm$   0.2 &  $-$ \\
P9 $\lambda$9229   &   2.25 $\pm$  0.07 &   2.4 $\pm$  0.4 & $-$ & $-$ & $-$ & $-$ \\
{[S III]} $\lambda$9532  &    7.8 $\pm$   0.4 &    8 $\pm$   1 &  $-$ &  $-$ &  $-$ &  $-$ \\
P8 $\lambda$9547  &    3.2 $\pm$   0.2 &  $-$ &  $-$ &  $-$ &  $-$ &  $-$ \\
E(B$-$V) &  0.08 $\pm$ 0.01  & 0.12  $\pm$ 0.01  & 0.07 $\pm$ 0.01  &   0.08 $\pm$0.01  & 0.24 $\pm$ 0.02 & 0.1 $\pm$ 0.1 \\
F(H$\beta$)[$10^{-15}$ erg/s/cm$^2$ ] &  3.3$\pm$0.1  &  1.6 $\pm$ 0.1   &  0.78$\pm$0.01   &  0.74 $\pm$ 0.01 &  1.11$\pm$0.07 &  0.14$\pm$0.05  \\
$v_{helio}$ [km/s] & 513$\pm$3 & 522$\pm$1 & 508$\pm$4 & 479$\pm$1  & 487$\pm$6 & 519$\pm$7 \\
\hline
\hline
 \end{tabular}
 \begin{tablenotes}
\small
\item Notes: Fluxes are given on a scale where $F(H\beta)=100$.  
\end{tablenotes}
 \end{table*}

\section{Temperatures, Densities and Chemical Abundances}

Temperatures, densities, and chemical abundances were obtained following the procedure described in \cite{annibali17}, using the  {\it getCrossTemDen}  and {\it getIonAbundance} tasks of the PyNeb code \citep{pyneb}. The {\it getCrossTemDen}  task  simultaneously derives   
electron densities ($n_e$) and temperatures ($T_e$)  through an iterative process  assuming a density-sensitive and a temperature-sensitive diagnostic line ratio.
We used the density-sensitive [S II] $\lambda6716/\lambda6731$ ratio to derive $n_e$ and, when possible, three sets of temperature-sensitive line ratios 
([O~III]$\lambda4363/\lambda4959+\lambda5007$, [S~III]$\lambda6312/\lambda9069$, and  [O~II]$\lambda7320+\lambda7330/\lambda\lambda3726,29$) to derive the temperature of the $O^{++}$,  $S^{++}$, and  $O^{+}$ ions, respectively. Table~\ref{tab_properties} summarises the results obtained for our \HII regions in DDO~68. 
The derived densities are typically low, with 
$n_e\lesssim100 \ cm^{-3}$, in agreement with the results found for other star-forming dwarf galaxies \citep[see, e.g.,][]{annibali15,annibali17}. 
The  $O^{++}$ temperature could only be determined for Reg.~1, Reg.~3, and  Reg.~7.
We found high T$_e (O^{++})$ in the range 17,000$-$19,000 K, in agreement with the results of \cite{Pustilnik05} and \cite{Izotov09}.  The $S^{++}$ temperature could only be obtained  for Reg.~1, due to the faintness of the [S~III]$\lambda$6312 line in the other regions; the derived value of $\sim$19,000 K  is in very good agreement with the theoretical relation 
between  $T_e(O^{++})$  and   $T_e(S^{++})$ proposed by \cite{garnett92} (see below). On the other hand, the $T_e(O^{+})$ obtained for Reg.~1 (and for Reg.~4) strongly deviates from the 
theoretical relations, a  problem that was already found and discussed by previous authors  \citep[e.g.][]{k03,bresolin09a,berg15} and that warns against the use of $O^{+}$ 
temperatures directly derived from the [O~II]$\lambda\lambda7320,30/\lambda\lambda3726,29$ ratio.

\begin{table*}
  \caption{Derived Properties for  \HII regions in DDO~68.}
  \label{tab_properties}
  \begin{tabular}{lcccc}
\hline
Line & Reg-1 & Reg-3 & Reg-7 & Reg-8 \\
 \hline      
 $n_e [cm^{-3}]$ &   $\leq$26 &   $76_ {-76}^{+169}$ &   $116_{-116}^{+148}$  &    $40_ {-2}^{+2}$\\
$T_e (O^+)$ [K] &   23000 $\pm$ 3000 &    $-$ &    $-$ &   13000 $\pm$  1000 \\
$T_e (O^{++})$ [K] &   19400 $\pm$ 400 &   18000 $\pm$ 2000 &   17000 $\pm$ 2000 &     $-$ \\
$T_e (S^{++})$ [K] &   19000 $\pm$  1000 &     $-$ &     $-$ &     $-$ \\
\hline
$(O^+/H^+)\times10^6$  &   1.66 $\pm$  0.08 &   5 $\pm$ 1 &   4 $\pm$ 1 &   $-$\\
$(O^{++}/H^+)\times10^6$  &  10.9 $\pm$  0.5 &   9 $\pm$ 2 &   5 $\pm$ 1 &   $-$\\
$12+ \log(O/H)$ &   7.10 $\pm$  0.02 &   7.14 $\pm$  0.07 &   6.96 $\pm$  0.09 &   $-$\\
\hline
$He/H$ &    0.082 $\pm$   0.003 &    0.079 $\pm$   0.002 &    0.085 $\pm$   0.010 &   $-$\\
$Y$ &    0.246 $\pm$   0.007 &    0.240 $\pm$   0.005 &    0.25 $\pm$   0.02 &   $-$\\
\hline
$(N^+/H^+)\times10^7$  &   0.7 $\pm$  0.1 &   1.8 $\pm$ 0.3 &  1.5 $\pm$ 0.1 &   $-$\\
$ICF(N^+)$  &  7.06 &  2.92 &  2.27 &  $-$ \\
$12+ \log(N/H)$ &   5.70 $\pm$  0.06 &   5.72 $\pm$  0.06 &   5.54 $\pm$  0.04 &   $-$\\
log(N/O)  &  -1.39 $\pm$  0.07 &  -1.4 $\pm$  0.1 &  -1.4 $\pm$ 0.1 &   $-$\\
\hline
$(Ne^{++}/H^+)\times10^6$  &   2.1 $\pm$  0.1 &   2.1 $\pm$ 0.4 &   1.3 $\pm$ 0.4 &   $-$\\
$ICF(Ne^{++})$  &  1.06 &  1.15 &  1.19 &  $-$ \\
$12+ \log(Ne/H)$ &   6.35 $\pm$  0.02 &   6.37 $\pm$  0.09 &   6.2 $\pm$  0.1 &   $-$\\
log(Ne/O)  &  -0.75 $\pm$  0.03 &  -0.8 $\pm$  0.1 &  -0.8 $\pm$  0.2 &   $-$\\
\hline
$(S^{+}/H^+)\times10^7$  &   0.31 $\pm$  0.02 &   0.8 $\pm$  0.1 &   0.9 $\pm$  0.1 &   $-$\\
$(S^{++}/H^+)\times10^7$  &   2.0 $\pm$  0.1 &   2.1 $\pm$  0.2 &   1.6 $\pm$  0.5 &   $-$\\
$ICF(S^++S^+{++})$  &  1.74 &  1.02 &  0.93 &  $-$ \\
$12+ \log(S/H)$ &   5.61 $\pm$  0.02 &   5.47 $\pm$  0.04 &   5.36 $\pm$  0.09 &   $-$\\
log(S/O)  &  -1.49 $\pm$  0.03 &  -1.67 $\pm$  0.08 &  -1.6 $\pm$  0.1 &   $-$\\
\hline
$(Ar^{++}/H^+)\times10^7$  &   0.47 $\pm$  0.03 &   0.59 $\pm$  0.08 &   0.4 $\pm$  0.1 &   $-$\\
$ICF(Ar^{++})$  &  1.26 &  1.08 &  1.07 &  $-$ \\
$12+ \log(Ar/H)$ &   4.77 $\pm$  0.03 &   4.80 $\pm$  0.06 &   4.7 $\pm$  0.1 &   $-$ \\
log(Ar/O)  &  -2.33 $\pm$  0.03 &  -2.34 $\pm$  0.09 &  -2.3 $\pm$ 0.1 &   $-$\\
\hline
 \hline
 \end{tabular}
 \end{table*}

Chemical abundances were derived for Reg.~1, Reg.~3, and Reg.~7 assuming a three-zone model for the electron temperature structure. 
The $T_e(O^{++})$ temperature was adopted for the highest-ionization zone ($O^{+2}$, $Ne^{+2}$, $He^{+}$, $He^{+2}$), the  $T_e(S^{++})$  temperature for the 
intermediate-ionization zone ($S^{+2}$, $Ar^{+2}$), and the  $T_e(O^{+})$ temperature for the low-ionization zone ($O^{+}$, $N^{+}$, $S^{+}$, $Fe^{+2}$).
For sake of homogeneity, and because of the aforementioned potential problems affecting  $T_e(O^{+})$ determinations, we computed the  $S^{++}$  and $O^{+}$  temperatures from the theoretical relations of \cite{garnett92}: 
 
 \begin{equation}
T_e[S~III] = 0.83 \times T_e [O~III] + 1700 \ K,
\end{equation}

\begin{equation}
T_e[O~II] = 0.70 \times T_e [O~III] + 3000 \ K.
\end{equation}

 To determine the abundances of the various ions, we used the extinction-corrected fluxes (listed in Table~\ref{tab_flux_corr}) of the following lines:  He~I~$\lambda$4471, 
He~I~$\lambda$5876,  and He~I~$\lambda$6678 for $He^{+}$, He~II~$\lambda$4686 (when available) for $He^{+2}$, [N II]$\lambda$6584 for $N^+$,  [O III]$\lambda\lambda$4959,5007 for $O^{+2}$, [O~II]$\lambda\lambda$3726,29 for  $O^{+}$,  [Ne~III]$\lambda$3869 for  $Ne^{+2}$, [S~II]$\lambda\lambda$6716,31 for  $S^{+}$,  [S~III]$\lambda9069$ for  $S^{+2}$, and [Ar~III]$\lambda$7136 for  $Ar^{+2}$.  
The PyNeb code adopts the He~I emissivities of \cite{porter12,porter13} including collisional excitation, so no correction to the emission line fluxes for this effect 
\citep{clegg87} needs to be applied. The He~I~$\lambda$7065 line, which has a strong contribution from collisional excitation, was not included in the 
computation of the  $He^{+}$; in fact, the uncertainties on the derived $n_e$ values translate into large errors on the  $He^{+}$ abundance due to the 
 strong dependence of the  He~I~$\lambda$7065 emissivity on density  \cite[see e.g. Fig.~4 of] []{porter12}. We did not use the faint HeI$\lambda$4922 line or the HeI$\lambda$5015 line, blended to the much stronger [O~III]$\lambda$5007 emission.
The ion abundance uncertainties were derived running the  {\it getIonAbundance} task for  ranges of temperatures, densities, and flux ratios within the $\pm 1 \sigma$ levels, and conservatively adopting the maximum excursion around the nominal abundance value as our error. 
In the case of $He^{+}$, the abundance was computed by averaging the individual abundances from the HeI$\lambda$4471, HeI$\lambda$5876,  and HeI$\lambda$6678 lines. 

Total element abundances were derived from the abundances of ions seen in the optical spectra using ionization correction factors (ICFs).
For Oxygen, the total abundance was computed as  $O/H= (O^+ + O^{+2})/H^+$; the contribution of $O^{+3}$ to the total oxygen abundance is in fact expected to  be  $<$1\%, since $O^+ /(O^+ + O^{+2}) >0.1$ in our \HII regions \citep{izotov06}. The He abundance was computed from $He^+$, with the exception of Reg.~1 where we added the contribution from  $He^{+2}$. 
On the other hand, the contribution from the broad He~II~$\lambda$4686  emission detected in Reg.~3 was not included, since likely of stellar origin. 
For all the elements other than O and He, we adopted the ICFs from  \cite{izotov06} for the low Z regime (Eq. (18) through (22) in that paper) to obtain total abundances.   

As mentioned in Section 2, the abundances of Reg.~1 and Reg.~3  (as well as those of Reg.~2 and Reg.~6, not present in our sample) were inferred with the direct $T_e$ method  
 by other authors \citep{Pustilnik05,Pustilnik08,izotov07a,Izotov09}.   
For these two regions, our abundances are consistent with theirs: more specifically, we obtain for  Reg.~1 an abundance of 12 + log(O/H)=7.10$\pm$0.02, within 1 $\sigma$  of the value of 7.14$\pm$0.03 from \cite{izotov07a} and within 2 $\sigma$  of the value of 7.19$\pm$0.05 from \cite{Pustilnik08}; 
for Reg.~3 we derive 12 + log(O/H)=7.14$\pm$0.07, within 1 $\sigma$ of the values of 7.15$\pm$0.04 and 7.10$\pm$0.06 from \cite{Izotov09} and \cite{Pustilnik08}, respectively. 
We therefore confirm  an extremely low  oxygen abundance in DDO~68' s \HII regions.
Furthermore, we report for the first time a 3-$\sigma$ direct method abundance measurement in a region in the ``tail'' of DDO~68, namely Reg.~7 
(since \cite{Pustilnik05} reported there a 1-$\sigma$ [O~III]$\lambda$4363 detection), finding an abundance of  12 + log(O/H)=6.96$\pm$0.09, even lower than in the ``head'' of the comet.
Our results are summarized in Table~\ref{tab_properties}.

\subsection{Indirect method} \label{section_indirect}

\begin{figure}
\includegraphics[width=\columnwidth]{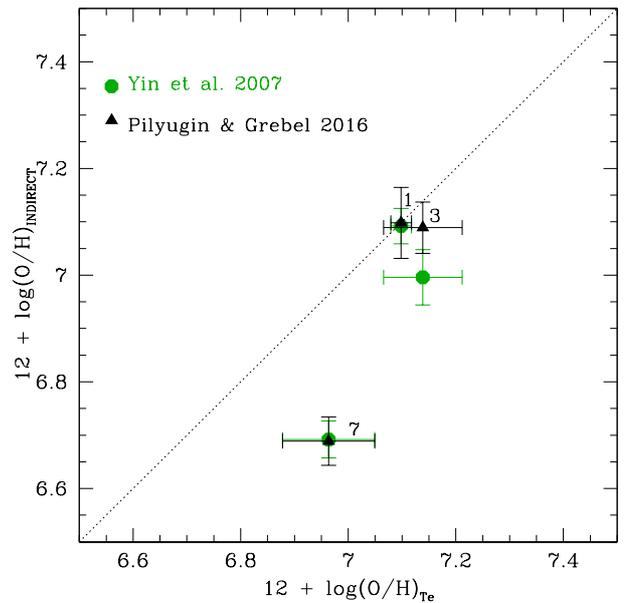}
 \caption{Oxygen abundances derived with the direct $T_e$ method and with different indirect-method calibrations \citep{yin07,pilyugin16} for Reg.~1, Reg.~3, and Reg.~7. The dotted line is the one-to-one relation. }
\label{indirect}
\end{figure}

Since [O~III]$\lambda$4363 was measured only in half of our targets, for all the \HII regions in our sample we also derived oxygen abundances using the indirect method.  As widely discussed in the literature, this method is affected by much larger uncertainties than the direct one, and furthermore different calibrations can disagree by as much as $\sim$0.5 dex \citep{kewley08}. 
For our study, we considered the two different calibrations proposed by 
\cite{yin07} and  \cite{pilyugin16}; 
we caution that these relations have been calibrated on direct-method abundances of 12 + log(O/H)$\gsim$7.10, whereas we are applying them to a lower metallicity regime for our study of DDO~68.
Figure~\ref{indirect} shows, for Reg.~1, Reg.~3, and Reg.~7 , the comparison between the abundances derived with the direct method and those obtained with the two different indirect-method calibrations. We notice that for Reg.~1 and Reg.~3 the Pilyugin \& Grebel calibration provides the best agreement 
with the direct-T$_e$ method. On the other hand, both calibrations underestimate by more than 0.2 dex the abundance derived for Reg.~7, the most metal-poor.
The results from the two calibrations for our six \HII regions are summarized in Table~\ref{tab_indirect}.

\begin{table}
  \caption{Oxygen abundances for  \HII regions in DDO~68 from direct and indirect methods.}
  \label{tab_indirect}
  \begin{tabular}{lccc}
\hline
\hline
Region & & 12 $+$ log(O/H) & \\
\hline
& $T_e$-Direct & Yin et al. 07 & Pilyugin \& Grebel 2016 \\
\hline
1 &  7.10 $\pm$  0.02 &    7.09 $\pm$ 0.02 &   7.10 $\pm$ 0.06 \\
3 &  7.14 $\pm$  0.07 &   7.00 $\pm$ 0.05   &  7.09  $\pm$ 0.04 \\
7 & 6.96 $\pm$  0.09 &     6.69 $\pm$ 0.03   &  6.69 $\pm$0.04\\
8 & $-$ & 7.28  $\pm$ 0.06 &  7.18 $\pm$ 0.04 \\
9 & $-$ & 7.0 $\pm$  0.3   & $-$ \\
4 & $-$ &  6.85  $\pm$ 0.01 &  6.83  $\pm$ 0.04 \\
 \hline         
\hline
 \end{tabular}
 \end{table}

\section{Abundances and abundance ratios of  H~II Regions}

As described in Section 2 and shown in Figure \ref{image}, Reg.~8 is close to the galaxy centre, Reg.~7 and Reg.~9 lie on the cometary tail that is often considered the accreted satellite of DDO~68, while Reg.~1, Reg.~3 and Reg.~4 design the ring-like structure which might be either the head of the comet/satellite or the periphery of the main body. We can thus try to use the abundances listed in Table~\ref{tab_properties} to get information on the evolutionary status of DDO~68, in terms either of a single galaxy or of two merged bodies.

\subsection{Helium}

Since the pioneering work by \cite{peimbert74},  the helium abundance derived from spectroscopy of the \HII regions in low-metallicity star-forming dwarfs has been considered as the best tool in the quest of the primordial helium abundance, and therefore an important test of the predictions of standard big bang nucleosynthesis \citep[SBBN, e.g.,][]{yang84}. 
 However, to make inferences on the SBBN, the primordial helium abundance must be inferred with an accuracy better than one percent, which is 
highly challenging, as widely discussed in the literature \citep[e.g.][]{olive04,izotov07b,aver11}. 
XMPs are extremely valuable for the determination of the primordial helium abundance, since the extrapolation of the He/H vs O/H distribution, or its mass fraction version $\Delta$Y/$\Delta$Z, to zero metallicity is minimal. For instance, from their samples of \HII regions in low-metallicity galaxies, \cite{izotov14} find Y$_P = 0.255 \pm 0.002$ (statistical + systematic error),  \cite{aver13} infer Y$_P = 0.2465 \pm 0.0097$, and \cite{fernandez18} find  Y$_P = 0.245 \pm 0.007$, marginally consistent with each other within the errors.

\begin{figure}
 \includegraphics[width=\columnwidth]{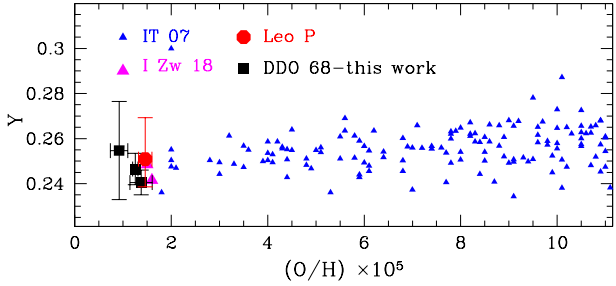}
 \caption{Helium mass fraction as a function of the oxygen abundance by number in DDO~68's \HII regions (black squares) compared with literature values for other metal-poor star-forming galaxies. The blue triangles are the low-metallicity \HII regions galaxies from \protect  \cite{izotov07b}, where we have highlighted the value inferred for IZw18 with the large magenta triangle; the large red circle corresponds to the helium mass fraction inferred by \protect \cite{skillman13} for Leo~P.}
\label{he}
\end{figure}

The extremely low oxygen abundance of DDO~68 makes it  a key object in the definition of the actual slope of He/H with O/H, that has been the subject of hot debates over decades \citep[see, e.g.,][and references therein]{izotov07b,peimbert07}. What is measured from the \HII region spectra is the abundance by number of helium relative to hydrogen, whereas the primordial helium abundance is usually estimated in mass fraction Y$_P$. We then convert our measured He/H values to the corresponding mass fraction Y via \cite{pagel92}'s simple relation:
\begin{equation}
 Y = $\{4 He/H [1-20(O/H)]\}$/$(1+4He/H)$.
 \end{equation}
 
 We infer helium mass fractions of Y=0.246$\pm$0.007, 0.240$\pm$0.005, and 0.25$\pm$0.02 for Regions 1, 3 and 7, respectively.
 Notice that the quoted errors  do not include systematic uncertainties due to the underlying theoretical assumptions and adopted physical parameters in the abundance computation. 
 \cite{peimbert07}, and references therein, evaluated the consequences on the derived primordial helium of several effects, including new atomic data. 
  \cite{izotov07b} extensively discussed all possible systematic effects that challenge high-accuracy determinations of helium abundances in H~II regions: He emissivities, reddening, temperature and ionization structure of the H~II region, underlying stellar He absorption, collisional excitation 
of hydrogen lines, deviation from case B recombination.
Using a Markov Chain Monte Carlo analysis, \cite{aver11} demonstrated that, even in presence of high-quality spectra of H~II regions, statistical and systematical uncertainties in the derived helium abundances can be as high as several percents.

Figure~\ref{he} shows how DDO~68's helium abundances derived here compare with the distribution of Y vs O/H in metal-poor star-forming galaxies. Our helium mass fractions lie at the low-oxygen end of the plot, and cover a range of values from slightly lower than in IZw18 \citep{izotov07b}, to slightly higher than in Leo~P \citep{skillman13}, with the warning that the highest value shows a significantly large error. 
 DDO~68's helium abundances are consistent with \cite{izotov14}'s, \cite{aver11}'s and \cite{fernandez18}'s  primordial values within 3 $\sigma$ uncertainties. We notice that \cite{fernandez18}  excluded from their linear regression the regions with higher He/H and N/O, arguing that they could be contaminated by nearby Wolf-Rayet stars. For this reason we do not attempt to derive a new regression line including also our data points, but simply show them in the plot.

Only in the last ten years or so, have people been able to compare the predictions of SBBN with other stringent constraints, resulting from the estimates of the baryon density from the satellites devoted to measuring the cosmic background radiation, WMAP \citep[Wilkinson Microwave Anisotropy Probe,][]{wmap} and the more recent Planck mission \citep{planck}.  
Under the assumption of SBBN and of a neutron lifetime of $\sim$880 s \citep{beringer12}, the baryon densities provided by the WMAP or the Planck experiments imply a primordial helium abundance of  Yp$=$0.2467$\pm$0.0006 \citep{spergel07,planck16}.  
 The He abundances determined by us for DDO~68 are therefore compatible with the primordial value from WMAP or Planck within the quoted uncertainties.

\protect \subsection{Spatial distribution of oxygen abundance}

\begin{figure}
\includegraphics[width=\columnwidth]{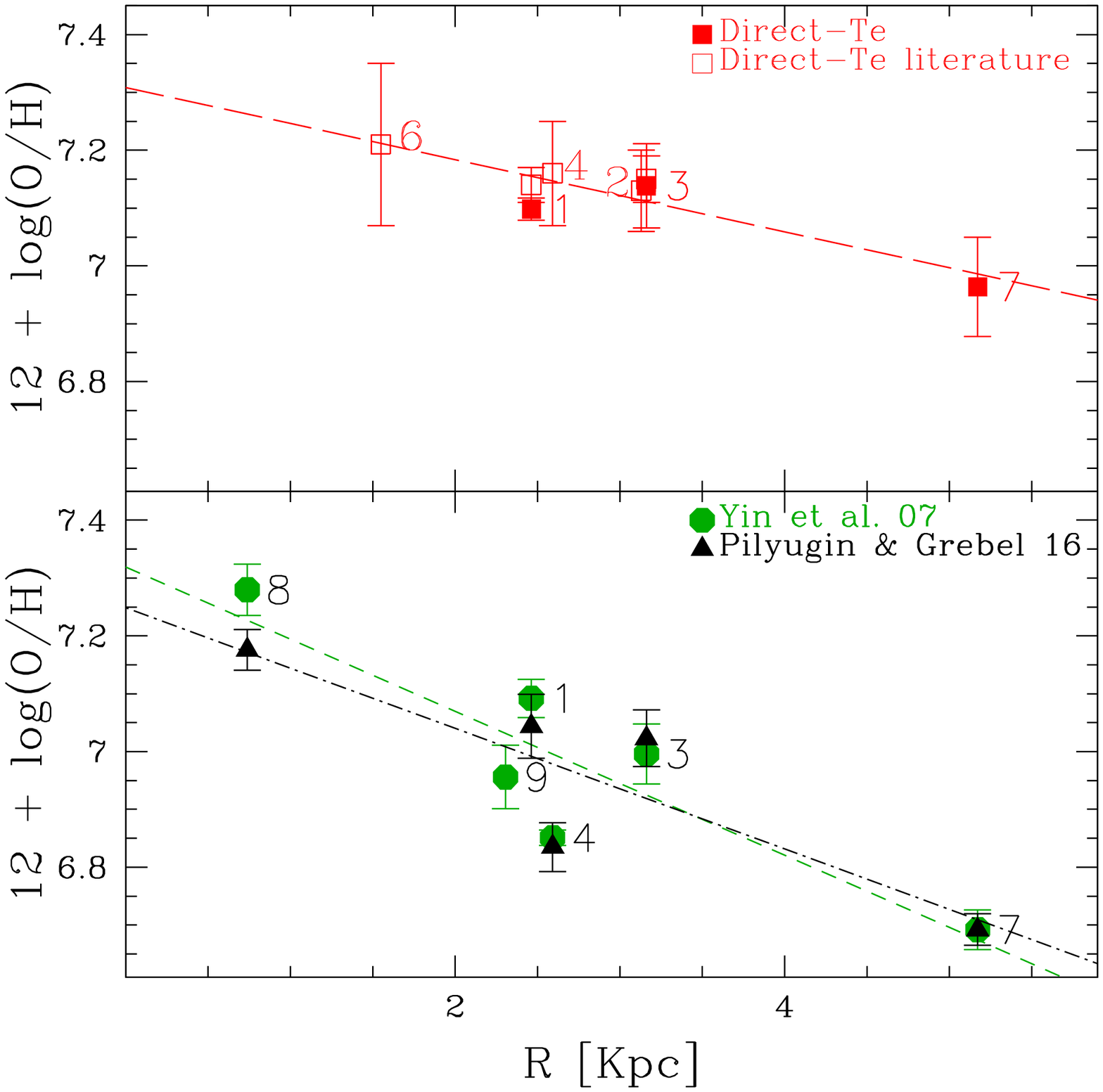}
 \caption{{\bf Top panel:} oxygen abundance as a function of the distance from the centre of DDO~68 derived with the direct T$_e$ method; full squares indicate our results, while open squares are the abundance values  from \protect \cite{izotov07a} and \protect \cite{Izotov09}. The dashed line is the linear fit to all values. 
 {\bf Bottom panel:} oxygen abundance as a function of the distance derived in this work  with the \protect \cite{yin07} and \protect  \cite{pilyugin16}'s indirect-method calibrations.
  The dashed and dot-dashed lines are the separate fits to the values obtained with the Yin et al. and Pilyugin \& Grebel calibrations, respectively. }
 \label{grad}
\end{figure}

We have analysed the spatial distribution of the oxygen abundances inferred for our \HII regions looking for signatures (or lack thereof) of the interactions with the galaxy's satellites. Table~\ref{tab_indirect} and Figure~\ref{grad} show that there is no clear evidence for the presence of a metallicity dichotomy in DDO~68, which could indicate a recent merging between two systems with very different initial metallicities. 
In particular, from an analysis of the color-magnitude diagram of the resolved stars, \cite{Tikhonov14} claimed the presence of two different metallicity components of 
$\sim$1/4 and $\sim$1/16  solar, belonging to the main system and to the accreted companion, respectively. However, from our \HII region analysis we do not find compelling evidence for two distinct metallicity populations in DDO~68, and in particular we do not detect a component as high in metallicity  as $\sim$1/4 solar (although there is no direct O abundance measurement in the most internal Reg.~8).

Rather than two separate metallicity populations, our analysis  suggests the presence of  a ``smooth''  trend of decreasing metallicity from the galaxy center outwards.
This is observed both for the direct-method abundance sample and for the indirect-method one.
In Figure~\ref{grad} the oxygen abundance has been plotted  
as a function of the distance from the galaxy centre, defined as the point of maximum light density (inferred from our HST and LBT imaging) and indicated by a cross in Figure~\ref{image}.  The top panel of Fig.~\ref{grad} shows the abundance results obtained in this work with the direct-T$_e$ method, complemented by direct-method abundances of DDO~68 from the literature \citep{izotov07a,Izotov09}. The linear fit to the combined sample provides a slope of $-0.06\pm0.03$ dex/kpc. 
In the bottom panel of Fig.~\ref{grad} we show instead the results obtained in this work with the \cite{yin07} and  \cite{pilyugin16}'s indirect-method calibrations (see Section~\ref{section_indirect}). The separate fits to the Yin et al. and Pilyugin \& Grebel  abundance sets provides slopes of  $-0.12\pm0.03$ dex/kpc and $-0.10\pm0.04$ dex/kpc, steeper than  the direct-method. 
Notice that the steeper gradients obtained with the indirect methods may be due to a poor calibration of the empirical relations at 
12 + log(O/H)$<$7.1, as already discussed in Section~\ref{section_indirect}.
In conclusion, we get evidence for a negative metallicity gradient in DDO~68, with $\Delta$log(O/H)/$\Delta$R in the range  $-0.06$ to  $-0.12$  dex kpc$^{-1}$,
 and no obvious chemical signature of a merging in our data.

In the past, metallicity gradients were considered quite small, or even absent, in dwarf galaxies \citep[see, e.g.,][and references therein]{skillman13}. However, until recently, very few, sometimes only one, \HII regions were usually observed in galaxies outside the Local Group.  Furthermore, uncertainties on the derived abundances could be quite large in absence of high signal-to-noise spectra. Gradients have been found in star-forming dwarfs when modern multi-object spectrographs have allowed people to measure  with significantly improved accuracy several \HII regions per galaxy  \citep{annibali15,pilyugin15,annibali17}, which suggests that others may have gone undetected for lack of appropriate data rather than for actual absence. 

For comparison with other literature studies it is useful to express the metallicity gradient in DDO~68 as a function of the isophotal radius R$_{25}$. Adopting R$_{25}=1.3^{\prime}$ from \cite{rc3} and D$=$12.7 Mpc from \cite{Sacchi16}, we obtain $R_{25}= 4.95$ kpc; 
the metallicity gradients are therefore $-0.30$  dex $R_{25}^{-1}$ (direct method) and $-0.59$  dex $R_{25}^{-1}$ (indirect method). For comparison, 
the metallicity gradients obtained with the direct method in the dwarfs NGC~4449 and NGC~1705 are $-0.29\pm0.08$ dex $R_{25}^{-1}$ \citep{annibali17} and  $-0.33\pm0.08$ dex $R_{25}^{-1}$ \citep{annibali15}, similar to the value obtained for DDO~68. 
Using a combination of direct and indirect-method abundances, \cite{pilyugin15} found gradients  in the range $-0.04\pm0.04$ to $-0.36\pm0.11$ dex $R_{25}^{-1}$ for a sample of 14 irregular galaxies, compatible with our direct-method gradient of DDO~68, but significantly flatter than the value obtained with the indirect method.  
Spiral galaxies display gradients too; using the indirect method calibration of \cite{Pilyugin12}, \cite{zinchenko16} found, for a sample of 88 disc galaxies in  the CALIFA survey, gradients in the range between -0.07 and -0.30  dex $R_{25}^{-1}$, with a few exceptions with gradient -0.64 dex $R_{25}^{-1}$. Thus, the metallicity gradient in DDO~68 is consistent with the steepest values derived  in spiral galaxies.

\subsection{Abundance Ratios}

\begin{figure*}
\includegraphics[width=19cm]{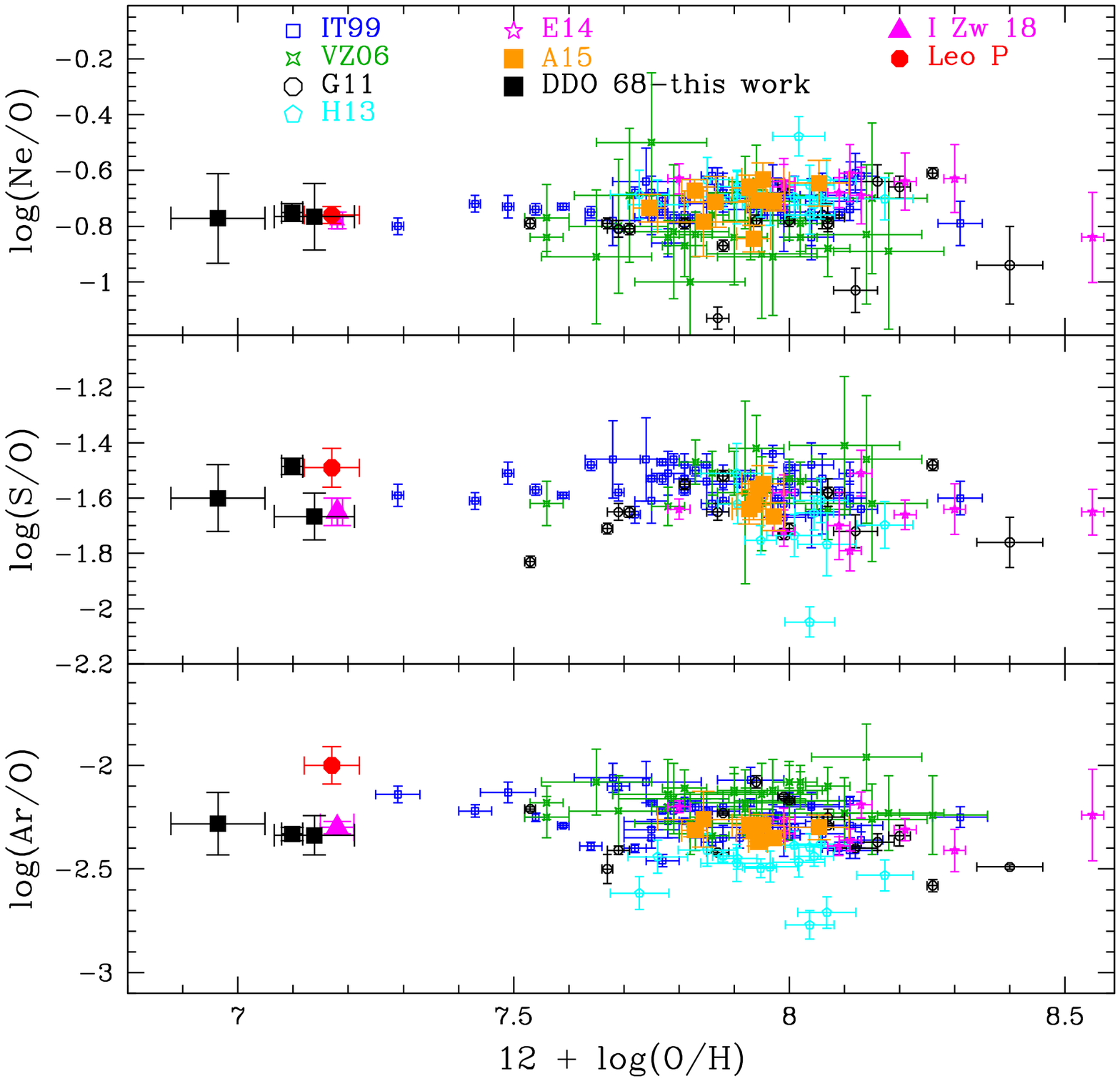}
 \caption{Chemical abundance ratio over oxygen of neon (top panel), sulfur (centre), and argon (bottom). The black filled squares show the ratios inferred for DDO~68 in this paper, while the other symbols show the literature values for other star forming dwarfs: IT99 \citep{IT99}; VZ06 \citep{VZ06}; G11 \citep{guseva11}; H13 \citep{haurberg13}; E14 \citep{esteban14}; A15 \citep{annibali15}. We have emphasized with larger symbols the ratios in IZw18 \protect \citep[magenta triangle;][]{IT99} and in LeoP \protect \citep[red circle;][]{skillman13}.}
\label{alpha-ratios}
\end{figure*}

Figure~\ref{alpha-ratios} shows the distribution of the abundance ratios Ne/O, S/O and Ar/O as a function of the oxygen abundance (direct method). 
The black squares correspond to the values we have derived for DDO~68's \HII regions, while the other symbols show the corresponding literature values for other metal-poor dwarfs. Neon, sulfur, and argon are all produced by the same massive stars that produce oxygen; hence, their abundance ratios to oxygen are expected to be independent of the oxygen abundance. This is indeed the case also in DDO~68, whose ratios lie all on the lower oxygen tail of the flat distribution inferred for the other dwarfs (see references in the caption to Fig.~\ref{alpha-ratios}). In this respect, although extremely oxygen-poor, DDO~68 behaves well, following the same average trend as the other galaxies.

\begin{figure}
 \includegraphics[width=\columnwidth]{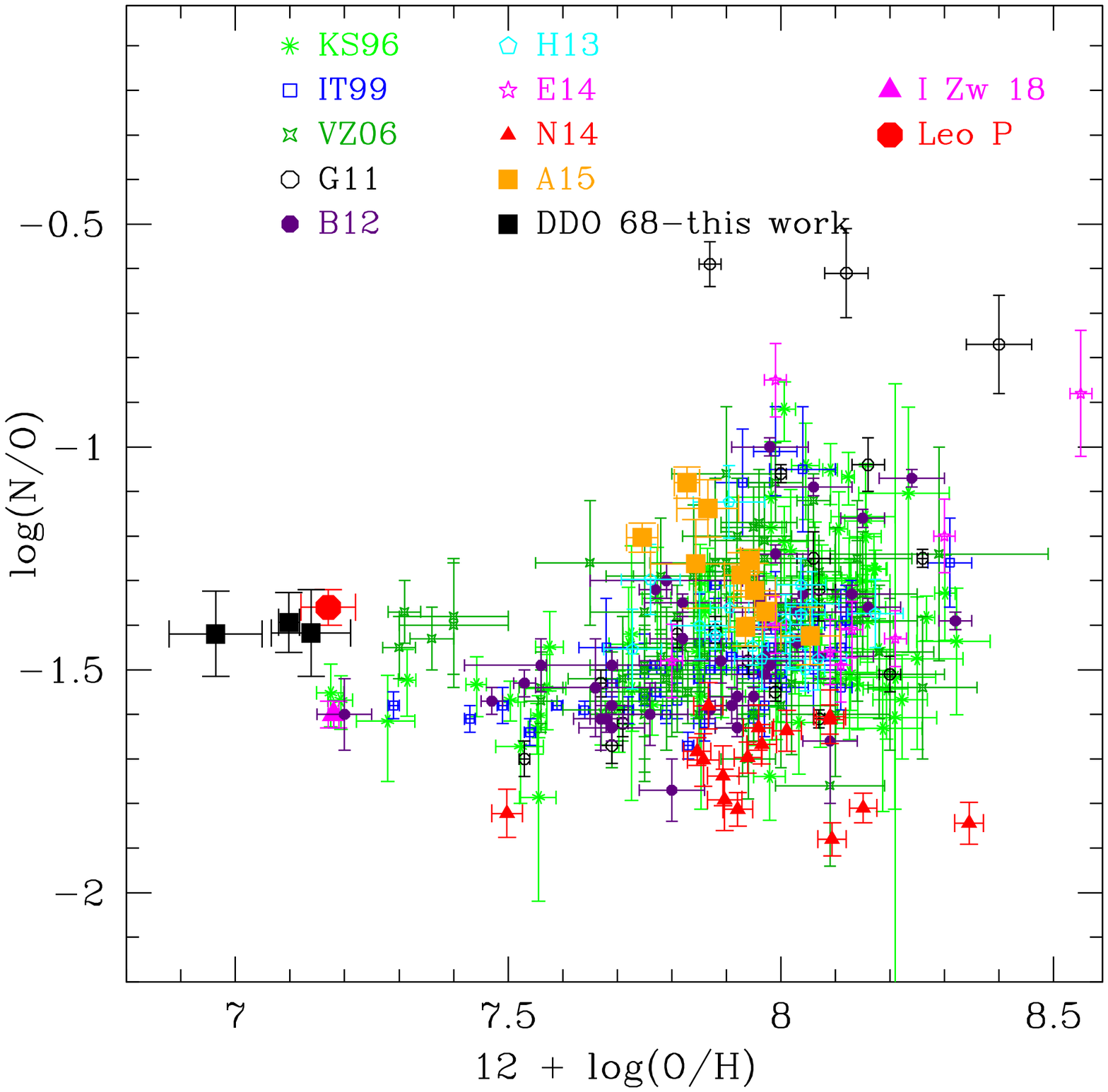}
 \caption{Nitrogen over oxygen abundance ratio as a function of the oxygen abundance. Symbols are as in Figure~\ref{alpha-ratios}.}
\label{n/o}
\end{figure}

Vice versa, DDO~68 does not appear to follow the general trend for what concerns nitrogen. Figure~\ref{n/o} shows the N/O ratio vs O/H in our \HII regions (black filled squares) and in those of the same large sample of star-forming dwarf galaxies plotted in Fig.~\ref{alpha-ratios}. While the very metal poor galaxy IZw~18, with its log(N/O)$\sim-1.6$, lies exactly on the tight plateau identified by \cite{IT99} for XMPs (i.e., galaxies with 12+log(O/H)$\leq$7.2), all the DDO~68' s \HII regions have even lower oxygen but significantly higher N/O, thus suggesting a hint of reverse trend with respect to that described by less metal-poor galaxies.

The N/O vs O/H diagram has been used to discuss the primary vs secondary nature of nitrogen for decades. A primary element is one (like oxygen) that can be synthesised in stars directly from primordial compositions, whereas a secondary element can be produced only in stars that at birth already contain seed/catalyst elements heavier than helium. As described by, e.g., \citet[][and references therein]{t80}, the canonical prediction for the abundances of secondary elements is that they increase as the square of primary abundances; hence, a secondary/primary abundance ratio is expected to scale linearly with the primary element abundance. The circumstance that N/O seems proportional to O/H in all galaxies with 12+log(O/H)$\geq$7.5$-$8.0
\citep[e.g.,][]{pagel85,vanzee98}, led people to conclude that nitrogen is a secondary element, a conclusion in agreement with standard stellar nucleosynthesis prescriptions that require catalyst carbon to produce N.
However, already forty years ago,  \cite{lequeux79} argued that the behaviour of the N/O ratios observed in a few irregular and blue compact galaxies should be ascribed to a non negligible fraction of primary nitrogen. 
Indeed, in 
1981, \cite{rv81} showed that nitrogen can be synthesised  as a primary element in favourable conditions, namely, when deep convection brings the seeds necessary for the nitrogen production to the appropriate stellar layer, even if they were not present in the original composition but were synthesised in the stellar core during earlier evolutionary phases. In practice, they demonstrated that, in the Hot Bottom Burning (HBB) evolutionary phase of intermediate mass stars, nitrogen is synthesised during their hydrogen-envelope burning, using as catalyst the carbon previously produced in the core of the same star. The existence of a primary (minority) fraction of N was confirmed by the combination of chemical evolution models with \HII region observations in the Milky Way and other spiral and late-type galaxies \citep[e.g.,][]{mt85,diaz86,vila93,marconi94}. In particular, it was shown that N/O is both observed and predicted to be rather flat across individual galaxies even in the oxygen-rich regime, although it increases with O/H from one galaxy to the other.

A further challenge to the interpretation of nitrogen's nature came from the N/O vs O/H diagram when people \citep[see, e.g.,][]{garnett90,ks96,IT99} suggested that in the lower oxygen regime log(N/O) shows a plateau. In particular, \cite{IT99} proposed that XMPs have a plateau at log(N/O)$\simeq-1.6$. Even invoking a significant fraction of primary nitrogen from HBB, this evidence was difficult to reconcile with standard nucleosynthesis predictions where oxygen is enriched much earlier than nitrogen because it is produced by massive, short-lived stars, while N should come from intermediate-mass, longer-lived ones.  As a way out of this issue, \cite{IT99} suggested that nitrogen could not only be primary but also be produced by low-metallicity massive stars, as partly envisaged by \cite{timmes95}. Indeed, \cite{maeder01} new stellar evolution models with rotation provided a theoretical support to this suggestion, since they predicted significant nitrogen enhancements in massive stars, with the N excesses being larger at larger stellar luminosities and at lower metallicity. 

Now, we might be facing a new challenge, since the log(N/O) we derived in all DDO~68's \HII regions is definitely higher than $-1.6$. The interesting point is that DDO~68 is not a lonely outlier, but shows almost exactly the same high ratio (red circle in Fig.~\ref{n/o}) inferred for Leo~P \citep{skillman13}, another XMP, and for few other low-metallicity systems \citep[see e.g.,][]{nava06,VZ06,izotov09b}. \cite{skillman13} discussed at length the apparent inconsistency of Leo~P's N/O with Izotov \& Thuan's plateau, posing key questions, but without reaching firm conclusions: is there actually a plateau or is it there a significant scatter (whether intrinsic or observational) as suggested by \cite{garnett90}? Are there systematic differences between emission line galaxies and relatively quiescent dwarfs? Does Leo~P show high N/O because it is observed during a long period of relative quiescence in which N enriches the medium but O does not, as in the time delay scenario proposed by \cite{garnett90}?

We do not have conclusions either, but we agree with them that in DDO~68, as in Leo~P, the current SF activity, although present, is much lower than in the XMPs that defined \cite{IT99} plateau (IZw18, SBS1415+437, SBS0335-052).  Its has been suggested \citep[e.g.,][]{pilyugin92,henry00} that the observed N/O ratio can be used as a ``clock'' to indicate the time since the most recent burst of star formation, due to the delayed release of N into the ISM compared to O. Since the highest activity of SF in DDO~68 has occurred between $\sim$10 and $\sim$300 Myr ago \citep{Sacchi16}, 
is it possible that the high N/O ratio measured there is due to a substantial nitrogen enrichment from the population of intermediate-mass stars, compared to a relatively lower oxygen enrichment from the less active recent SF?

Or, given the strong dependence on rotational velocity and metallicity of the nitrogen primary production in massive stars \citep{maeder01}, could it be that in these XMPs N is actually produced in stars more massive than the O-producers, and is thus released to the surrounding medium earlier than O? In other words, could we invoke a reversed time-delay model in XMPs, with N/O decreasing for increasing O/H? The most conservative option would be to attribute the high N/O (and the high He/H) to the presence of Wolf-Rayet stars close to our \HII regions, following the argument by \cite{fernandez18}. Indeed, a broad He~II$\lambda$4686 feature, possibly due to the presence of WR stars, is detected in Reg.3.

\section{Radial velocities}

Radial velocities were derived from the most prominent emission lines
in the spectra. We derived mean velocities averaging the H$\delta$,
H$\gamma$, H$\beta$ and [O~III]$\lambda\lambda$4959,5007 lines in the
blue MODS channel, and the He~I$\lambda$5876 and H$\alpha$ lines in
the red channel. The operation was performed separately for the MODS1
and MODS2 spectra, for a total of four velocity estimates for each
region.  The velocity measurements turned out to be more accurate 
(as quantified by the standard deviation) for the
red channel than for the blue one, and for MODS2 than for MODS1. 
The red and blue channel velocities typically agreed within the uncertainties.

The final velocities were then computed as the weighted mean
of the four different values, and we adopted the weighted standard
deviation as our velocity uncertainty. The derived velocities were
transformed into heliocentric ones using the {\it rvcorrect} task in
IRAF. The results are provided at the end of
Table~\ref{tab_flux_corr}. The velocities derived for our six \HII
regions range from 479$\pm$1 km/s to 522$\pm$1 km/s.  For comparison,
systemic  velocities for DDO~68 derived in the literature from neutral and ionized gas are in the
range $V_{sys}\sim$500$-$510 km/s \citep{Ekta08,Cannon14,moiseev14}.

It is useful to compare our results with the velocity map of the
ionized gas derived by \cite{moiseev14} for DDO~68 (see their
Fig. 1). \cite{moiseev14} found a velocity gradient in the north-south
direction, with $V-V_{sys}$ ranging from $\sim$-40 km/s to $\sim$+40
km/s from the ``tail'' to the ``head'' of the comet. This behaviour is
consistent with the rotation observed in the HI
\citep{Ekta08,Cannon14}, and indicates that  neutral and ionized gas rotate in
the same direction. Our velocity measurements are consistent with
this trend since the most external Reg.~7 exhibits the lowest radial
velocity (479$\pm$1 km/s), whereas the highest velocity (522$\pm$1
km/s) is found in Reg.~3 in the head of the comet. For comparison, the
HI velocities as inferred from the velocity contours displayed in
Fig. 2 of \cite{Cannon14} are $\sim$480 km/s for Reg.~7 and $\sim$530
km/s for Reg.~1 and 3, consistent with our values. For Reg.~4, the HI
velocity is $\sim$525 km/s, while we find a lower velocity of
508$\pm$4 km/s.  For Reg.~9, the HI velocity is $\sim$510 km/s versus
our \HII region value of 519$\pm$7 km/s, while for Reg.~8, the innermost 
one, the HI velocity is $\sim$505 km/s versus our lower value
of 487$\pm$6 km/s.  Therefore we conclude that, on a first
approximation, the  ionized gas kinematics trace the behaviour of the HI;
small deviations from the HI behaviour are nonetheless present, which
is not surprising since \HII region peculiar velocities, caused by
e.g.\ expanding shells, can significantly distort the disk velocity
field due to the relatively small amplitude of DDO~68 's rotation
curve \citep[$V_c=45$ km/s at 11 kpc radius,][]{Cannon14}. Indeed,
\cite{Pustilnik17} suggested that the arrangement of \HII regions in a
ring-like structure in the head of the comet is due to the feedback
effect from a supergiant shell expanding at a velocity of $\sim$13
km/s.
 
As mentioned in the introduction, \cite{anni16} presented an N-body
simulation of three interacting galaxies that, under a given line of
sight, is able to reproduce the morphology of the tail, the stream and the arc in DDO~68.
With the new information from our derived radial velocities, combined with literature studies of the ionized and 
HI gas, we will be able to perform new better-constrained dynamical simulations of the 
DDO~68 system.

\section{Discussion and Conclusions}
 
In this paper we have presented LBT/MODS spectroscopy of six \HII regions in DDO~68, two candidate companion galaxies, and one candidate star cluster. Our target \HII regions cover all the main components of the DDO~68 system (namely: main body, cometary tail and northern ring), except the small stream discovered by \cite{anni16}, whose objects are too faint for spectroscopy even with the LBT.
Both the galaxies and the candidate cluster turned out to be background galaxies.

We confirm the extremely low metallicity of DDO~68 and furthermore provide the first direct-method abundance measurement in the ``tail'' (Reg.~7); here we derive a metallicity as low as 12+log(O/H)$=$6.96$\pm$0.09, potentially lower, within the uncertainties, than that of the record-holder galaxy J0811+4730, whose oxygen abundance is 12+log(O/H)$=$6.98$\pm$0.02 \citep{izotov18}.

\begin{figure}
 \includegraphics[width=\columnwidth]{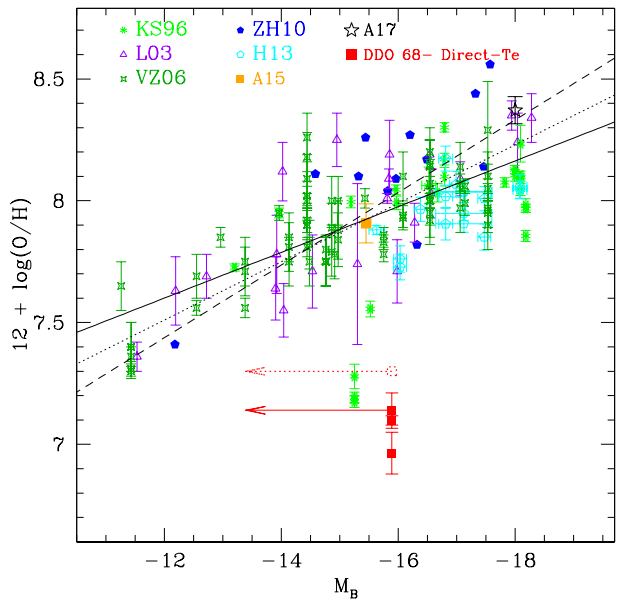}
 \caption{ 
  Luminosity - metallicity  (L-Z)  relation in star-forming dwarfs from the literature. All abundances have been obtained with the direct T$_e$ method. 
 Red filled squares indicate the abundances derived by us for Reg.1, 3 and 7 in DDO~68 with the  direct T$_e$ method. The dotted square at 
 $12 + log(O/H)\sim7.3$ is the value extrapolated at R$=$0 from the linear fit in Fig.~\ref{grad}. The arrows indicate the position of  DDO~68 in the  L-Z relation assuming that the measured H~II regions originated from a secondary acquired body with mass 1/10 that of DDO~68. The other symbols correspond to the labelled references: KS -\protect \cite{ks96}, L03 - \protect \cite{lee03}, VZ06 - \protect \cite{VZ06}, ZH10 - \protect \cite{zhao10}, H13 - \protect \cite{haurberg13},  A15 - \protect \cite{annibali15}, A17 - \protect \cite{annibali17}.
The solid line is the 3-$\sigma$ rejection linear least squares fit to the data points, excluding DDO~68: $12 + log(O H)=(-0.094\pm0.001) \times M_B + (6.47\pm0.12)$. The dashed and dotted lines are, respectively, the relations from \protect \cite{VZ06} and \protect \cite{haurberg13}.} 
\label{lumox}
\end{figure}

As a consequence, we confirm that DDO~68 is an extreme outlier in the luminosity-metallicity relation. Figure~\ref{lumox} shows the relation as inferred from the data by various authors and our DDO~68 values for Reg.~1, 3 and 7,  all obtained with the direct T$_e$ method; DDO~68's  absolute B magnitude is $M_B=-15.9$, computed adopting B=14.69 \citep{taylor05}, a foreground extinction of A$_B=0.08$ \citep{reddening}, and a distance modulus of  $(m-M)_0$=30.52 \citep{Sacchi16}. 
The average oxygen abundance inferred for DDO~68' s H~II regions is a factor of $\sim$ten lower than expected for its luminosity. 
We also show in the figure a hypothetical central abundance for DDO~68 obtained by extrapolating  the linear fit in Fig.~\ref{grad} to R$=$0; even this value is significantly below the  luminosity-metallicity relation, by a factor of about 5. 
If we attributed the \HII regions to a secondary acquired body with mass, and hence luminosity, ten times lower that DDO~68, and thus shifted them to the left, they would still remain outside the plot area occupied by the other galaxies (see arrows in Fig.~\ref{lumox}), with only the extrapolated central value being marginally consistent with the relation. 

From synthetic CMDs, \cite{Sacchi16} derived a total stellar mass of $\sim10^8 \ M_{\odot}$ for DDO~68, which allows us to explore the position of this galaxy on the mass-metallicity (M-Z) relation; taking as reference the distribution of galaxies in Fig.~8 of \cite{berg12}, we conclude that DDO~68 is an outlier also in the M-Z relation. 
It is worth recalling that the other known XMPs do not deviate from the overall relations. Also Leo~P \citep{skillman13}, which is somewhat similar to DDO~68 for  what concerns the low metallicity and the high N/O, is perfectly aligned with the other dwarfs \citep{berg12}, thanks to its much lower luminosity (and mass).  

The overall interpretation of the luminosity - metallicity relation is that late-type dwarfs with higher gas fractions and lower masses are more inefficient at chemically enriching their medium, both because they have proportionally larger amounts of gas to pollute and because they have more chances to lose their metals through galactic winds, due to their lower potential well \citep[e.g.,][]{Chisholm18}. This scenario can still apply also to DDO~68, but the question is why has its enrichment been so less efficient than the others. How did DDO~68 components manage to get rid of most of the oxygen produced by their stars? Or to dilute it with exceptional amounts of primordial external gas? In the past, some \citep{IT99} suggested that XMPs are experiencing now their first burst of star formation, and have not had time to significantly enrich and mix their medium. However, we now know that this is not the case for any of the dwarfs studied with deep HST photometry in the local Universe \citep[see, e.g.,][and references therein]{tht09,mcquinn10,weisz11}, all of which clearly contain stars as old as the lookback time reachable at their distance. In particular, DDO~68, as well as IZw18 and all the other XMPs within $\sim$18 Mpc, hosts also RGB stars, i.e. stars at least 2-3 Gyr, possibly 13 Gyr old \citep{aloisi07,Sacchi16}. The SFHs derived from our HST photometry by \cite{Sacchi16} and, independently, by \cite{Makarov17} agree that the system has been forming stars since the oldest epochs, although at a quite low rate until a few hundred Myr ago; hence oxygen must have been synthesized. 

Thanks to the space distribution of our targets, we have been able to detect a dependence of the oxygen abundance on the distance from the galaxy center, whose linear fit slope $\Delta$log(O/H)/$\Delta$R is in the range $-0.06$ to  $-0.12$  dex kpc$^{-1}$, or $-0.30$ to $-0.59$  dex $R_{25}^{-1}$,  with the flatter values from the direct T$_e$ method, and the steepest ones from the indirect one.
Unfortunately, the sample is too small to assess whether this dependence is actually a metallicity gradient similar to those observed in all spiral galaxies \citep[see, e.g.,][and references therein]{wilson94,Sanchez14} and, recently, in some star-forming dwarfs \citep{annibali15,pilyugin15,annibali17}, or the result of a yet incomplete mixing of DDO~68's two merging bodies, the main body and its satellite cometary tail (with or without the head in the northern ring). A combination of the two explanations is also possible, since born-in-situ gradients are the consequence of the spatial dependence of the ratio between star formation history (i.e. element production, usually higher in the inner regions) and infall of metal-poor gas (i.e. element dilution, usually higher in the outer regions), with the possible complication of galactic outflows (element loss, where the potential well is sufficiently low). At any rate, the \HII regions  in the tail have oxygen abundances lower than most of the others, hence the oxygen gradient observed in DDO~68 may be the consequence of the accretion of both metal-poor gas  and the metal-poorer satellite.

In a forthcoming paper (Romano et al. in preparation), we will address the evolution of the system with chemical evolution models taking into account the known SFH, the possible interaction histories of the system and the N-body simulations presented by \cite{anni16}, to try to disentangle whether the anomalously low observed oxygen is more likely due to the interactions, to unusually high accretion of primordial gas, or to galactic winds. 
We will furthermore investigate the reason for the high N/O observed in DDO~68,  trying to disentangle whether it can be explained with nitrogen enrichment from massive, low-metallicity rotating stars, as predicted by stellar models \citep[e.g.][]{maeder01,meynet02}, or if the reason is the age of the strongest burst occurred in DDO~68 (from $\sim$10 Myr to $\sim$300 Myr ago), combined with the delayed production of nitrogen with respect to oxygen \citep{henry00}.

Finally, we measured radial velocities in the range 479$\pm$1 km/s to 522$\pm$1 km/s from the tail to the head of the ``comet'' in DDO~68, consistent with the rotation derived in the H~I by \cite{Cannon14}.

\section*{Acknowledgments}

 We thank the anonymous referee for his/her highly valuable comments and suggestions and for his/her very accurate report, which contributed to significantly improve the paper.
F.A., D.R. and M.T. kindly acknowledge funding from INAF PRIN-SKA-2017 program 1.05.01.88.04.
This work was based on LBT/MODS data. 
The LBT is an international collaboration among institutions in the United States, Italy and Germany. LBT Corporation partners are: the University of Arizona on behalf of the Arizona Board of Regents; Istituto Nazionale di Astrofisica, Italy; LBT Beteiligungsgesellschaft, Germany, representing the Max-Planck Society, the Leibniz Institute for Astrophysics Potsdam, and Heidelberg University; the Ohio State University, and the Research Corporation, on behalf of the University of Notre Dame, University of Minnesota and University of Virginia.

\appendix{}

\section{LBT/MODS spectra of target regions.}

\begin{figure*}
\includegraphics[width=\textwidth]{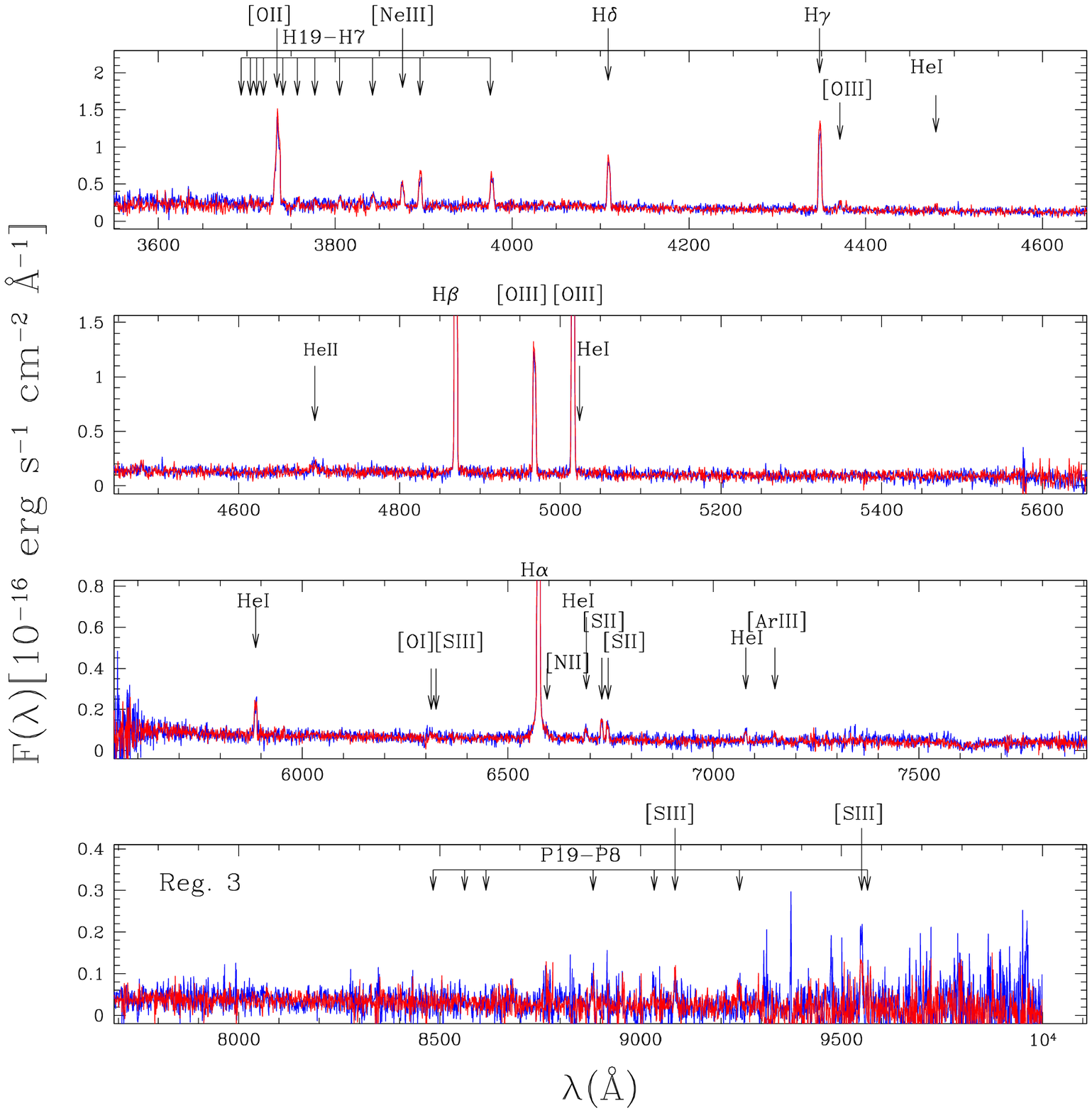}
 \caption{LBT/MODS spectra for Reg.~3  in  DDO~68 acquired with MODS1 (blue line) and MODS2 (red line). Indicated are the most relevant emission lines. }
\label{spectra_reg3}
\end{figure*}

\begin{figure*}
\includegraphics[width=\textwidth]{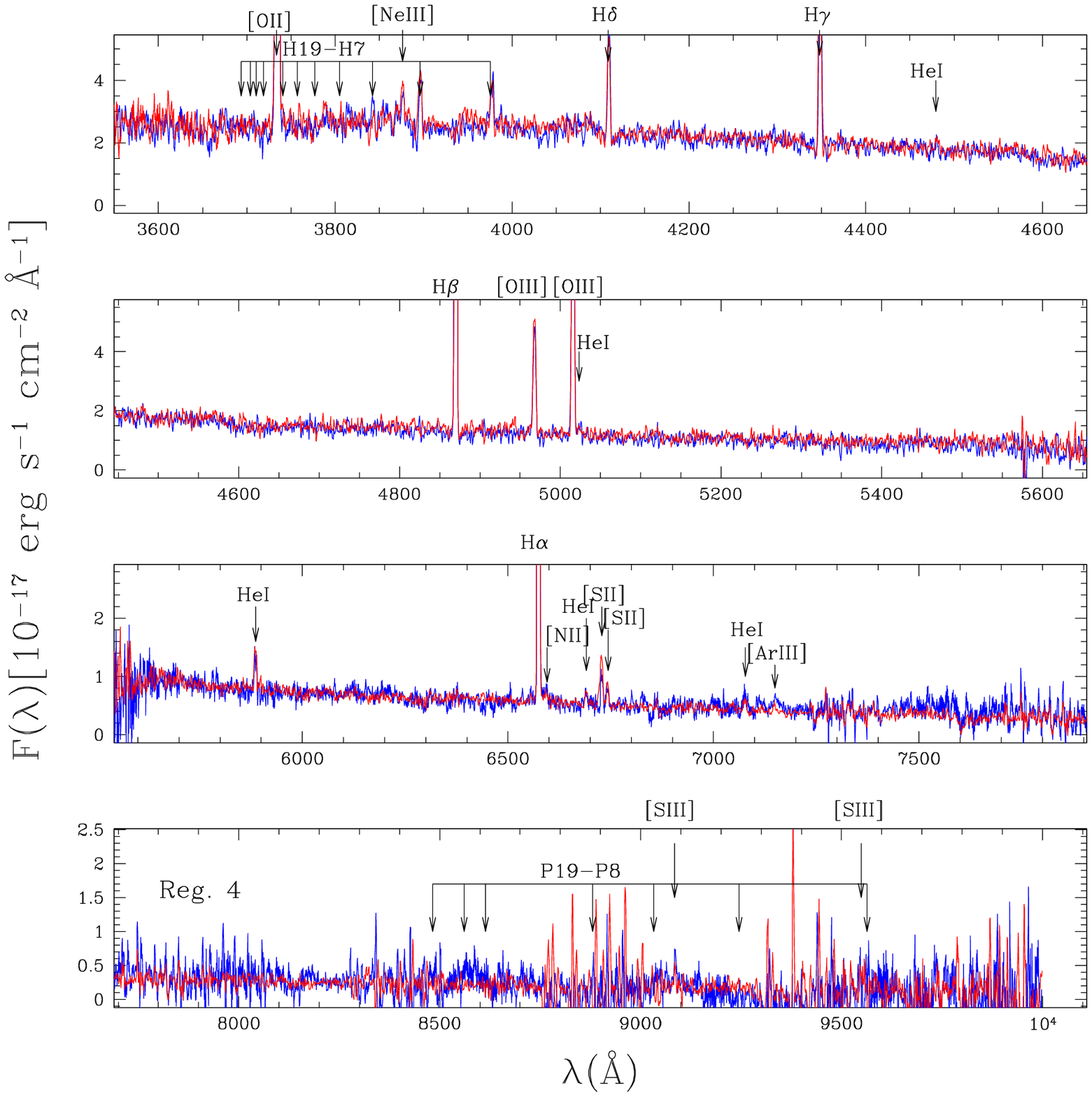}
 \caption{LBT/MODS spectra for Reg.~4  in  DDO~68 acquired with MODS1 (blue line) and MODS2 (red line). Indicated are the most relevant emission lines.}
\label{spectra_reg4}
\end{figure*}

\begin{figure*}
\includegraphics[width=\textwidth]{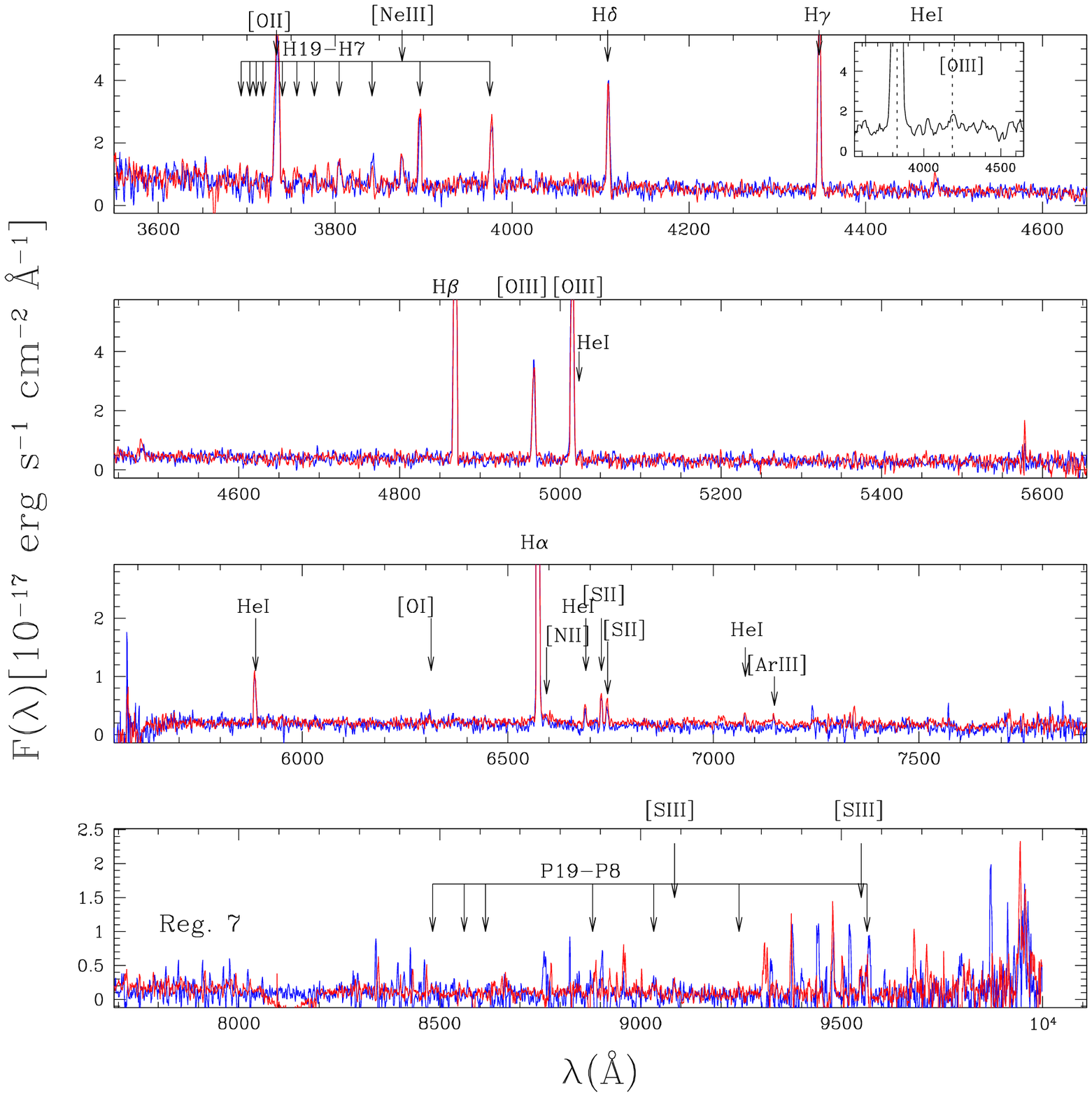}
 \caption{LBT/MODS spectra for Reg.~7  in  DDO~68 acquired with MODS1 (blue line) and MODS2 (red line). Indicated are the most relevant emission lines. The small insertion in the top panel shows the spectrum obtained from the combination of all MODS1 and MODS2 data, where the [O III]$\lambda$4363 detection is visible.}
\label{spectra_reg7}
\end{figure*}

\begin{figure*}
\includegraphics[width=\textwidth]{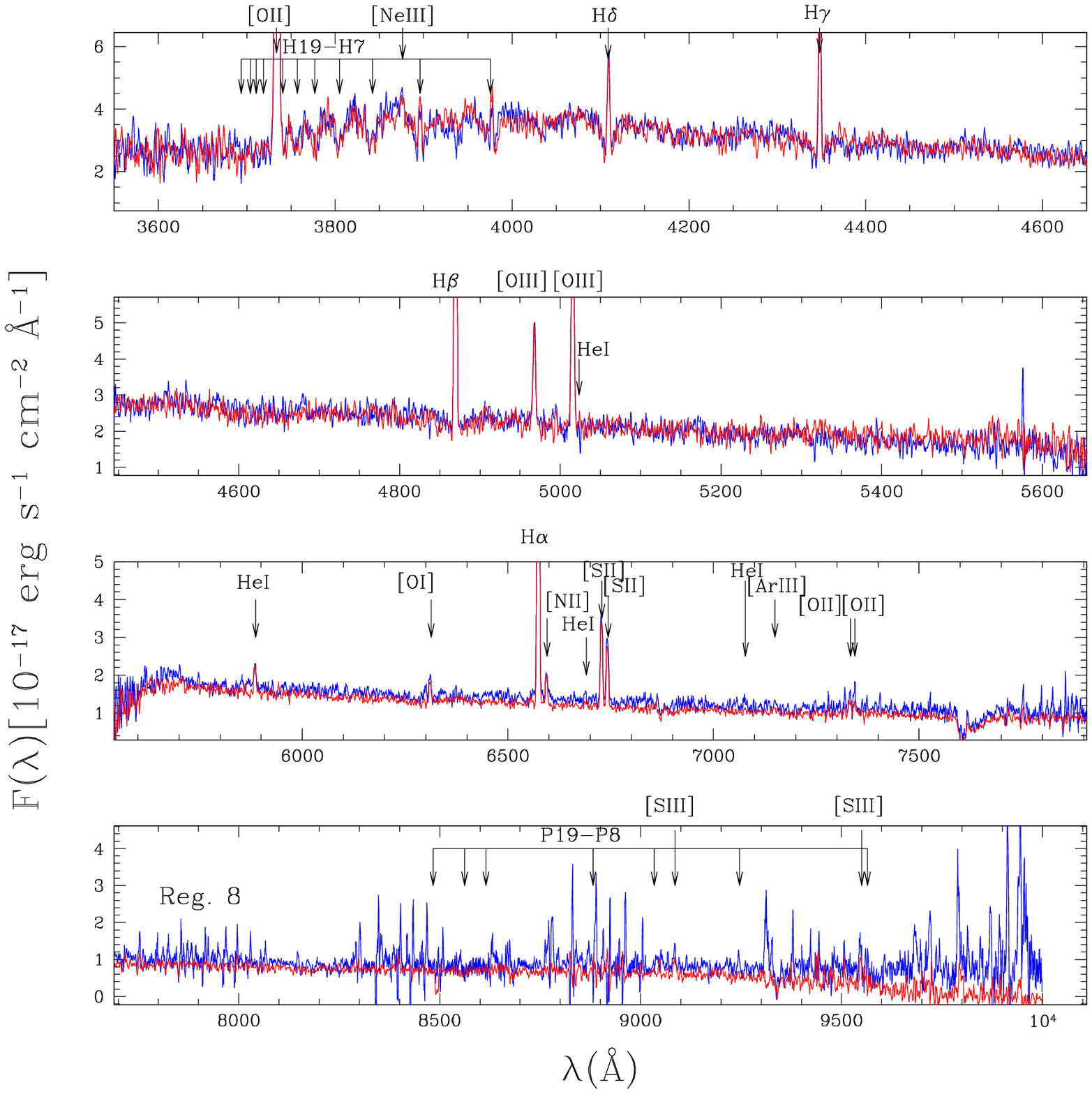}
 \caption{LBT/MODS spectra for Reg.~8  in  DDO~68 acquired with MODS1 (blue line) and MODS2 (red line). Indicated are the most relevant emission lines.}
\label{spectra_reg8}
\end{figure*}

\begin{figure*}
\includegraphics[width=\textwidth]{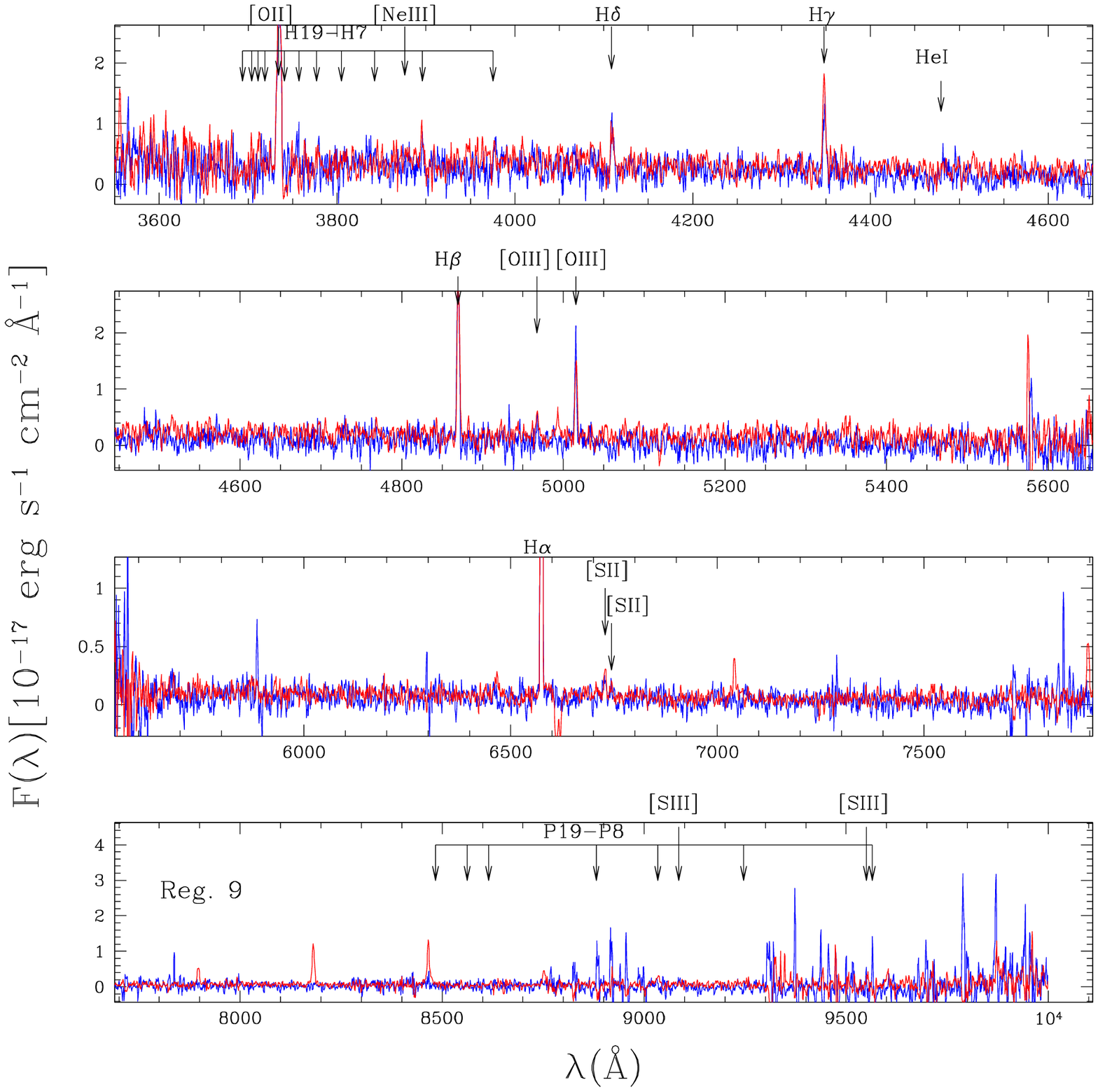}
 \caption{LBT/MODS spectra for Reg.~9  in  DDO~68 acquired with MODS1 (blue line) and MODS2 (red line). Indicated are the most relevant emission lines.}
\label{spectra_reg9}
\end{figure*}

\begin{figure*}
\includegraphics[width=\textwidth]{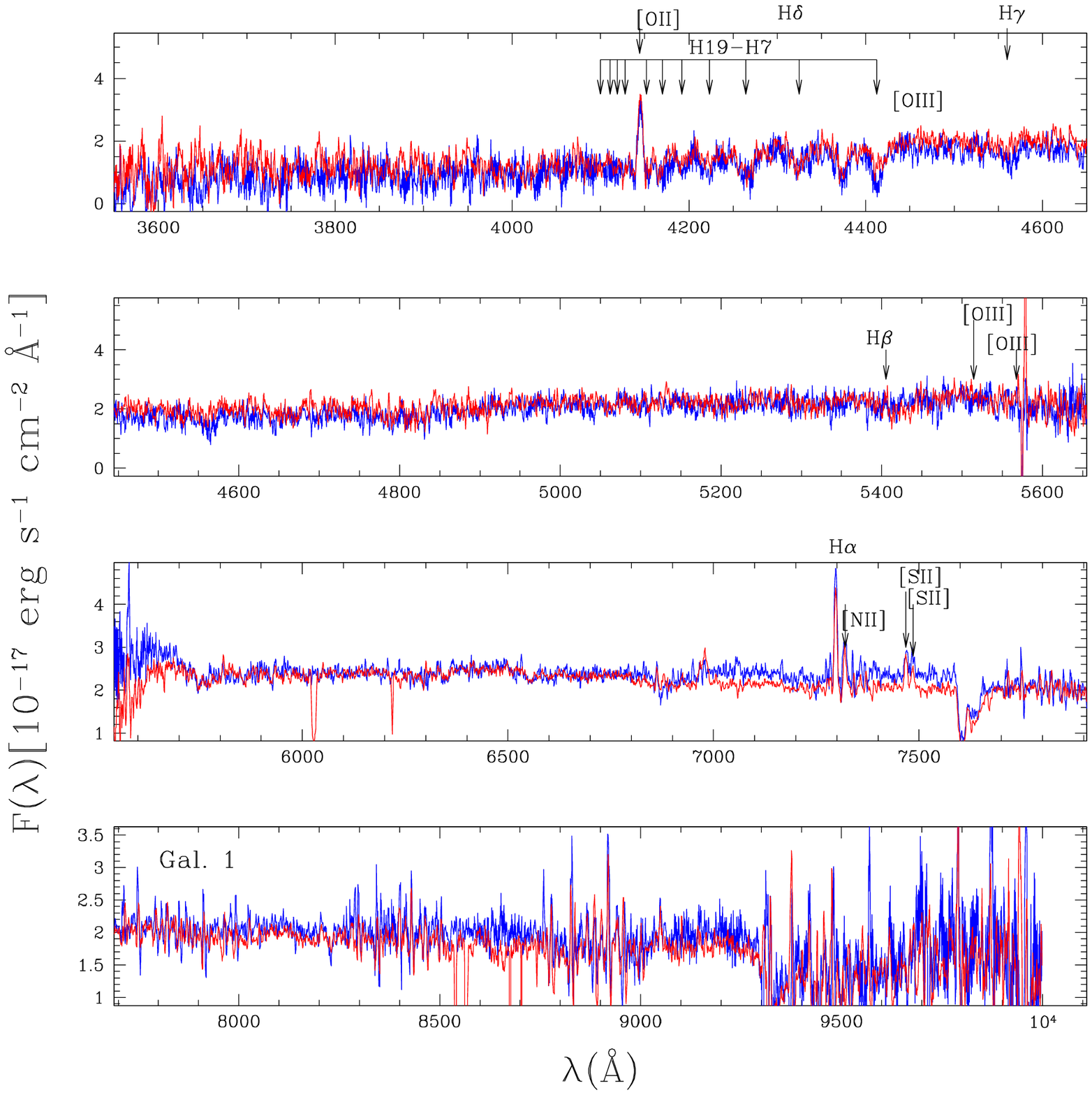}
 \caption{LBT/MODS spectra for Gal.~1, one of the two candidate companion galaxies of DO~68, acquired with MODS1 (blue line) and MODS2 (red line). Indicated are the most relevant emission and absorption lines. The system turned out to be an emission-line galaxy at z$\sim$0.11.}
\label{spectra_gal1}
\end{figure*}

\begin{figure*}
\includegraphics[width=\textwidth]{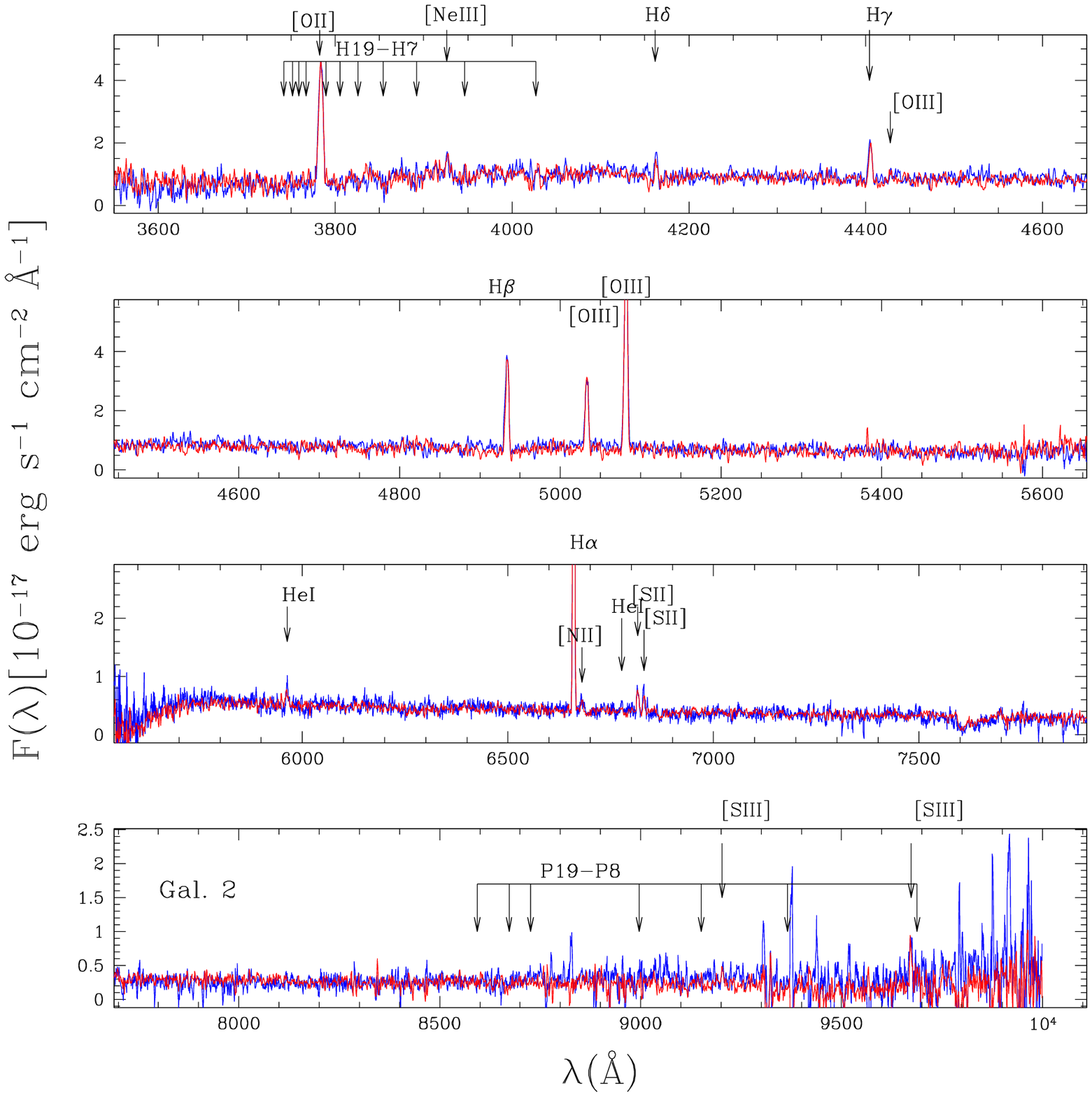}
 \caption{LBT/MODS spectra for Gal.~2, one of the two candidate companion galaxies of DO~68, acquired with MODS1 (blue line) and MODS2 (red line). Indicated are the most relevant  emission lines. The system turned out to be an emission-line galaxy at z$\sim$0.0148.}
\label{spectra_gal2}
\end{figure*}

\begin{figure*}
\includegraphics[width=\textwidth]{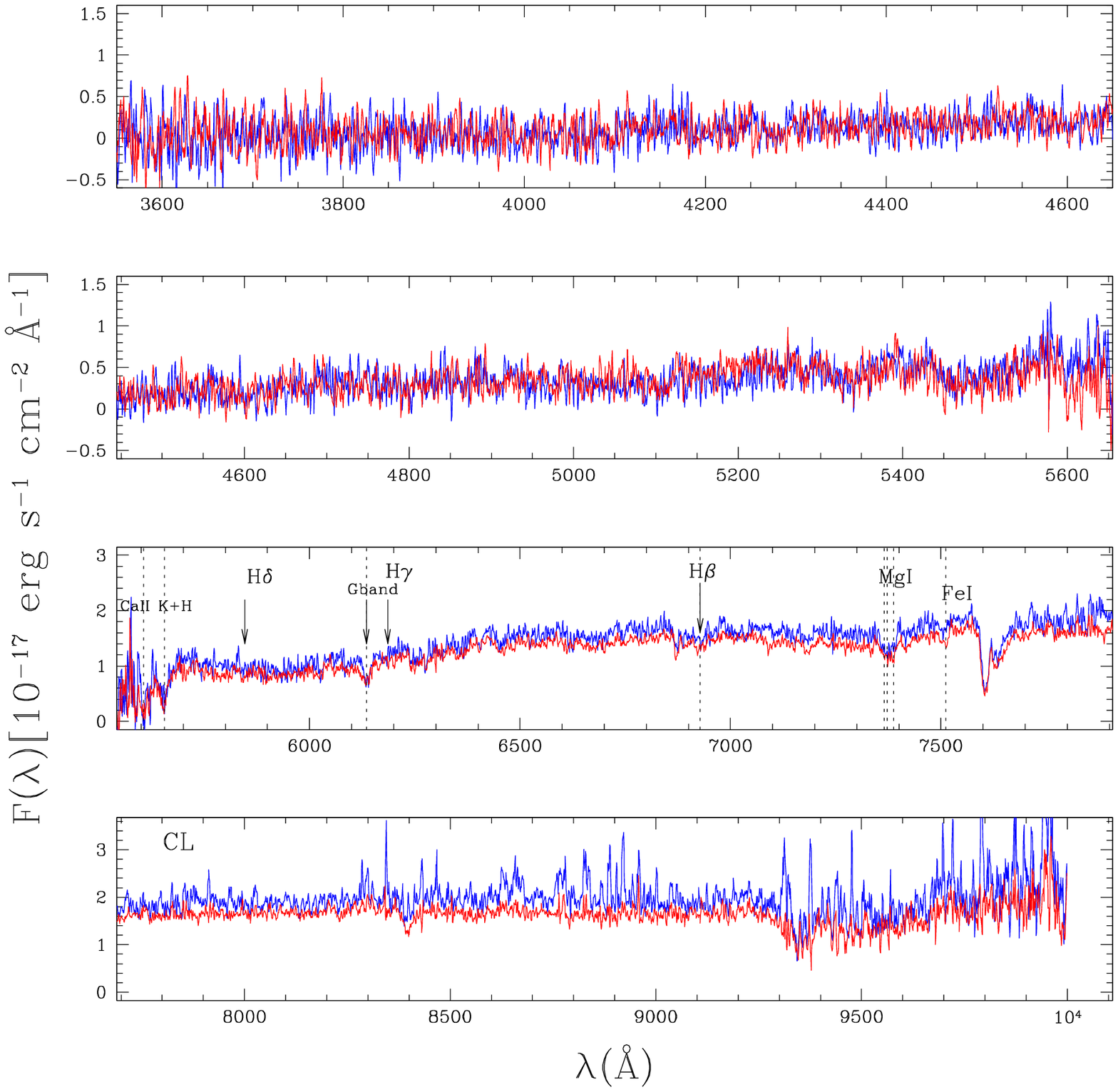}
 \caption{LBT/MODS spectra  for CL, a candidate star cluster in DO~68, acquired with MODS1 (blue line) and MODS2 (red line).  The system turned out to be a background early-type galaxy at z$\sim$0.42.}
\label{spectra_cl}
\end{figure*}

The LBT/MODS spectra of \HII regions 3, 4, 7, 8 and 9 in DDO~68, of the two candidate companion galaxies (Gal.~1 and Gal.~2), and of the candidate star cluster (CL) are shown in Figures~\ref{spectra_reg3} to ~\ref{spectra_cl}.

\section{``Raw'' emission line fluxes}

We provide in Tables~\ref{tab_flux_raw} the emission line fluxes, with no reddening correction applied, for the six \HII regions studied in DDO~68. The reported fluxes were obtained by averaging the results from MODS1 and MODS2, as outlined in Section~\ref{section_fluxes}, with the associated uncertainties computed as the standard deviation of the two different measurements. 

 \begin{table*}
  \caption{Observed emission fluxes for \HII regions in DDO~68.}
  \label{tab_flux_raw}
  \begin{tabular}{lcccccc}
\hline
Line & Reg-1 & Reg-3 & Reg-4 & Reg-7 & Reg-8 & Reg-9 \\
 \hline         
{[O II]} $\lambda$3727 &     6.20 $\pm$    0.09 &     6.0 $\pm$    0.3 &     4.8 $\pm$    0.1 &     2.6 $\pm$    0.4 &    11.2 $\pm$    0.6 &    1.48 $\pm$   0.06 \\
H10 $\lambda$3978 &     1.06 $\pm$    0.01 &     0.39 $\pm$    0.05 &   $-$ &     0.31 $\pm$    0.03 &   $-$ &   $-$ \\
He I $\lambda$3820  &  $-$ &    0.27 $\pm$   0.01 &  $-$ &  $-$ &  $-$ &  $-$ \\
H9$+$He II $\lambda$3835 &    1.44 $\pm$   0.06 &    0.56 $\pm$   0.05 &  $-$ &    0.3 $\pm$   0.1 &  $-$ &  $-$ \\
{[Ne III]} $\lambda$3869 &     3.66 $\pm$    0.06 &     1.18 $\pm$    0.05 &     0.58 $\pm$    0.08 &     0.38 $\pm$    0.01 &   $-$ &   $-$ \\
H8$+$He I $\lambda$3889 &     4.4 $\pm$    0.2 &     1.7 $\pm$    0.2 &     0.76 $\pm$    0.04 &     0.9 $\pm$    0.1 &     0.72 $\pm$    0.08 &   $-$ \\
H$\epsilon$ $+$ He I $+$[Ne III] $\lambda$3970 &     4.51 $\pm$    0.08 &     1.73 $\pm$    0.09 &     0.67 $\pm$    0.06 &     0.88 $\pm$    0.04 &     0.64 $\pm$    0.07 &   $-$ \\
H$\delta$ $\lambda$4101 &     6.1 $\pm$    0.1 &     2.68 $\pm$    0.07 &     1.28 $\pm$    0.01 &     1.34 $\pm$    0.01 &     1.1$\pm$    0.1 &     0.32 $\pm$    0.02 \\
H$\gamma$ $\lambda$4340 &    11.5 $\pm$    0.4 &     4.8 $\pm$    0.3 &     2.61 $\pm$    0.03 &     2.54 $\pm$    0.03 &     2.01 $\pm$    0.03 &     0.51 $\pm$    0.08 \\
{[O III]} $\lambda$4363 &    1.42 $\pm$   0.04 &    0.35 $\pm$   0.06 &    $-$ &    0.094 $\pm$   0.005 &   $-$ &  $-$ \\
He I $\lambda$4471 &     0.78 $\pm$    0.08 &   $-$ &     0.15 $\pm$    0.01 &     0.23 $\pm$    0.01 &   $-$ &   $-$ \\
He II (WR) $\lambda$4686 &  $-$ &    0.9 $\pm$   0.2 &  $-$ &  $-$ &  $-$ &  $-$ \\
He II  $\lambda$4686 &    0.577 $\pm$   0.001 &  $-$ &  $-$ &  $-$ &  $-$ &  $-$ \\
H$\beta$ $\lambda$4861 &    25.3 $\pm$    0.8 &    10.9 $\pm$    0.6 &     6.01 $\pm$    0.05 &     5.65 $\pm$    0.07 &     5.0 $\pm$    0.2 &     1.06 $\pm$    0.01 \\
He I $\lambda$4922 &     0.26 $\pm$    0.02 &   $-$ &   $-$ &   $-$ &   $-$ &   $-$ \\
{[O III]} $\lambda$4959 &   16.1 $\pm$   0.5 &    4.7 $\pm$   0.2 &    1.57 $\pm$   0.05 &    1.35 $\pm$   0.02 &    1.03 $\pm$   0.05 &    0.14 $\pm$   0.01 \\
{[O III]} $\lambda$5007 &    46.5 $\pm$    1.5 &    13.8 $\pm$    0.7 &     4.5 $\pm$    0.1 &     3.81 $\pm$    0.04 &     3.2 $\pm$    0.2 &     0.54 $\pm$    0.07 \\
He I $\lambda$5015  &    0.59 $\pm$   0.03 &  $-$ &  $-$ &  $-$ &  $-$ &  $-$ \\
He I $\lambda$5876  &     2.44 $\pm$    0.05 &     1.12 $\pm$    0.01 &     0.47 $\pm$    0.01 &     0.56 $\pm$    0.01 &     0.5 $\pm$    0.1 &   $-$ \\
{[OI]}  $\lambda$6302 &    0.135 $\pm$   0.009 &  $-$ &  $-$ &    0.08 $\pm$   0.02 &    0.39 $\pm$   0.02 &  $-$ \\
{[S III]} $\lambda$6312 &    0.18 $\pm$   0.01 &  $-$ &  $-$ &  $-$ &  $-$ &  $-$ \\
H$\alpha$ $\lambda$6563  &    77 $\pm$   2 &    35.3 $\pm$    0.6 &    18.5 $\pm$    0.1 &    17.3 $\pm$    0.2 &    17.2 $\pm$    0.1 &     3.5 $\pm$    0.1 \\
{[N II]} $\lambda$6584  &     0.29 $\pm$    0.04 &     0.29 $\pm$    0.01 &     0.18 $\pm$    0.02 &     0.12 $\pm$    0.01 &     0.51 $\pm$    0.02 &  $-$  \\
He I $\lambda$6678   &    0.72 $\pm$   0.03 &    0.324 $\pm$   0.005 &    0.170 $\pm$   0.007 &    0.18 $\pm$   0.01 &    0.13 $\pm$   0.02 &  $-$ \\
{[S II]} $\lambda$6716  &     0.59 $\pm$    0.01 &     0.59 $\pm$    0.07 &     0.5 $\pm$    0.2 &     0.30 $\pm$    0.01 &     1.58 $\pm$    0.01 &     0.16 $\pm$    0.01 \\
{[S II]} $\lambda$6731  &     0.40 $\pm$    0.02 &     0.44 $\pm$    0.01 &     0.20 $\pm$    0.05 &     0.23 $\pm$    0.02 &     1.14 $\pm$    0.05 &     0.07 $\pm$    0.01 \\
He I $\lambda$7065  &    0.645 $\pm$   0.001 &    0.32 $\pm$   0.02 &    0.16 $\pm$   0.01 &    0.111 $\pm$   0.008 &  $-$ &  $-$ \\
{[Ar III]} $\lambda$7136  &    0.49 $\pm$   0.01 &    0.242 $\pm$   0.009 &    0.102 $\pm$   0.006 &    0.08 $\pm$   0.03 &  $-$ &  $-$ \\
He I $\lambda$7281  &    0.118 $\pm$   0.006 &  $-$ &  $-$ &  $-$ &  $-$ &  $-$ \\
{[O II]} $\lambda$7320  &    0.19 $\pm$   0.02 &  $-$ &  $-$ &  $-$ &    0.23 $\pm$   0.02 &  $-$ \\
{[O II]} $\lambda$7330  &    0.15 $\pm$   0.03 &  $-$ &  $-$ &  $-$ &    0.27 $\pm$   0.04 &  $-$ \\
P10 $\lambda$9017  &    0.484 $\pm$   0.003 &  $-$ &  $-$ &  $-$ &  $-$ &  $-$ \\
{[S III]} $\lambda$9069  &    1.03 $\pm$   0.06 &    0.45 $\pm$   0.03 &    0.20 $\pm$   0.07 &    0.16 $\pm$   0.05 &    0.28 $\pm$   0.01 &  $-$ \\
P9 $\lambda$9229   &    0.669 $\pm$   0.003 &    0.34 $\pm$   0.05 &  $-$ &  $-$ &  $-$ &  $-$ \\
{[S III]} $\lambda$9532  &    2.3 $\pm$   0.1 &    1.1 $\pm$   0.2 &  $-$ &  $-$ &  $-$ &  $-$ \\
P8 $\lambda$9547  &    0.95 $\pm$   0.04 &  $-$ &  $-$ &  $-$ &  $-$ &  $-$ \\
\hline
\hline
 \end{tabular}
 \begin{tablenotes}
\small
\item The fluxes, in units of $10^{-16} erg s^{-1} cm^{-2}$ \AA$^{-1}$, are not corrected for reddening or for Balmer/Helium line absorption. 
    \end{tablenotes}
 \end{table*}




\newpage








\bsp	
\label{lastpage}
\end{document}